\documentclass[twocolumn]{aastex63}

\usepackage{amsmath}
\usepackage{cleveref}

\received{April 29, 2020}
\revised{June 30, 2020}
\accepted{July 13, 2020}

\shortauthors{Boocock, Kusano and Tsiklauri}

\begin{document}

\title{The Effects of Oscillations \& Collisions of Emerging Bipolar Regions on the Triggering of Solar Flares}

\author{C. Boocock}
\affiliation{School of Physics and Astronomy,
Queen Mary University of London,
Mile End Road, London, E1 4NS,
Great Britain}

\author[0000-0002-6814-6810]{K. Kusano}
\affiliation{Solar-Terrestrial Environment Laboratory, Nagoya University, Furo-cho, Chikusa-ku, Nagoya, Aichi 464-8601, Japan }

\author[0000-0001-9180-4773]{D. Tsiklauri}
\affiliation{School of Physics and Astronomy,
Queen Mary University of London,
Mile End Road, London, E1 4NS,
Great Britain}

\begin{abstract}

The ability to predict the occurrence of solar flares in advance is important to humankind due to the potential damage they can cause to Earth's environment and infrastructure. It has been shown in \citet{Kusano_2012} that a small-scale bipolar region (BR), with its flux reversed relative to the potential component of the overlying field, appearing near the polarity inversion line (PIL) is sufficient to effectively trigger a solar flare. In this study we perform further 3D magnetohydrodynamic simulations to study the effect that the motion of these small-scale BRs has on the effectiveness of flare triggering. The effect of two small-scale BRs colliding is also simulated. The results indicate that the strength of the triggered flare is dependent on how much of the overlying field is disrupted by the BR. Simulations of linear oscillations of the BR showed that oscillations along the PIL increase the flare strength whilst oscillations across the PIL  detract from the flare strength. The flare strength is affected more by larger amplitude oscillations but is relatively insensitive to the frequency of oscillations. In the most extreme case the peak kinetic energy of the flare increased more than threefold compared to a non-oscillating BR. Simulations of torsional oscillations of the BR showed a very small effect on the flare strength. Finally, simulations of colliding BRs showed the generation of much stronger flares as the flares triggered by each individual BR coalesce. These results show that significantly stronger flares can result from motion of the BR along the PIL of a sheared field or from the presence of multiple BRs in the same region.

\end{abstract}


\section{Introduction}

Solar and stellar flares are of importance to humankind due their
hazardous effect on life on surrounding planets. In the case of the Sun
and Earth, \citet{Eastwood_2017} discuss in details 
the economic impact of space weather, including the
solar flares. \citet{Baumen_2014} estimate that
for a 1989 Quebec-like event, the global economic impacts would range from $2.4$  to $3.4$ trillion dollars over a year. 
Statistics of stellar super-flares is discussed in \citet{Shibayama_2013}.
They find that, in the case of the Sun-like stars (with surface temperature 5600-6000 K and 
slowly rotating with a period longer than 10 days), the occurrence rate of super-flares with energies of 
$10^{34}-10^{35}$ erg is once in 800-5000 yr.

It is believed that a physical process that adequately describes solar flares is magnetic reconnection -- 
a rapid change of connectivity of magnetic field lines, during which
magnetic energy is converted into other forms of energy such as heating (thermal energy increase) 
and kinetic energy (KE) of plasma outflows, if a continuum based description such as magnetohydrodynamics (MHD)
is used or super-thermal particles, or if kinetic (particle-based) plasma description
is used. While the broad brush picture of solar flares is well accepted \citep{Masuda_1994,Shibata_1995},
the details of the flare triggering mechanism is not. \citet{Kusano_2012} shed light
on the topic of flare triggering. As summarized by  \citet{Kusano_2012} previous
 works have considered relationships between the occurrence 
 of solar eruptions and different magnetic properties, such as
 (i) strong magnetic shear,
 (ii) reversed magnetic shear,
 (iii) sigmoidal structure of the coronal magnetic field,
 (iv) flux cancellation (FC) on the photosphere,
 (v) converging foot point motion,
 (vi) the sharp gradient of the magnetic field, 
 (vii) emerging magnetic fluxes, 
 (viii) presence of multipolar topologies,
 (ix) the presence of flux ropes, 
 (x) narrow magnetic lanes between major sunspots, and possibly many others.
More recently, in order to elucidate the stability problem of the pre-eruptive state, \citet{Ishiguro_Kusano_2017}
developed a simple model in which the sigmoidal field is modelled by a double arc electric 
current loop and its stability is analyzed. As a result, they 
found that the double arc loop is more easily destabilized than the axisymmetric torus, 
and it becomes unstable even if the external field does not 
decay with altitude, which is in contrast with the axisymmetric torus instability. 
In order to understand the flare trigger mechanism, \citet{2017ApJ...842...86M} 
conducted three-dimensional magnetohydrodynamic simulations using a 
coronal magnetic field model derived from data observed by the Hinode satellite. 
Their aim was to investigate what kind of magnetic disturbance may trigger the flare. 
As a result,  \citet{2017ApJ...842...86M} found that certain small bipole fields,
 which emerge into the highly sheared global magnetic field of an active region, 
 can effectively trigger a flare. These bipole fields can be classified into 
 two groups based on their orientation relative to the polarity inversion line (PIL): 
 the so-called opposite polarity and reversed shear structures, 
 as suggested by \citet{Kusano_2012}.

The main findings of the pioneering work by \citet{Kusano_2012} were made by
systematically surveying the nonlinear dynamics caused by a wide variety of magnetic 
structures in terms of three-dimensional magnetohydrodynamic simulations. 
As a result, they determined that two different types of small-scale BRs
favor the onset of solar eruptions. These BRs, which should 
appear near the magnetic PIL, include magnetic 
fluxes reversed to the potential component or the nonpotential component 
of the major field on the PIL. The central finding of \citet{Kusano_2012}
was illustrated in their Figure 2, where they considered maximum KE achieved
during a solar eruption as a function of shear angle $\theta_0$ and 
the injected small-scale field  azimuthal orientation $\phi_e$.
It was found that large values of $\theta_0$ close to $90^\circ$ and $\phi_e=180^\circ$
produce the most favourable conditions for the solar flares.

In this work we consider the most favourable condition for the solar flares
to be the same as that in \citet{Kusano_2012}, but now, in addition,
we impose a linearly polarized
oscillation on the emerging small-scale BR. We study
the effect of amplitude and frequency variation of linearly polarized oscillations of single, emerging BRs 
and the collision of two emerging BRs on solar flare efficiency.
Our motivation is twofold: 

\begin{enumerate}
  \item We would like to investigate how oscillations in the BR, both linear and torsional, might affect the previous results of \citet{Kusano_2012}. Such oscillations can come from Alfv\'en waves travelling along the emerging magnetic flux tube that extends from solar corona down to below the  photosphere.

  \item Magnetic features such as sunspots are known to collide/coalesce. Hence we would like to study how the collision of two BRs affects the amount of energy released during the flare modelled by emerging, and at the same time colliding, small-scale fields interacting with an overlaying, preexisting field with $\theta_0=80^\circ$ and $\phi_e=180^\circ$ configuration.
\end{enumerate}

We find that movement of the small-scale BRs along the PIL increases the strength of the flare triggered whilst movement of the BR away from the PIL decreases the flare strength. Torsional motion seems to have little effect on the flares; though sustained rotation that moves the BR too far from its most favourable orientation ($180^\circ$ relative to the PIL) does reduce the strength of the flare. Finally collisions lead to more energetic flares due to both the movement of the individual BR and the coalescence of the flares triggered by each emerging BR.

Section 2 gives details of the simulation setup.
Section 3 provides the main results of this study. 
Section 4 provides a discussion of the results and
Section 5  closes this work with a list of conclusions.  

\section{Simulation Setup}

In order to investigate the effect of dynamics in the emerging BR, simulations were performed in Lare3d (\citet{lare}). The setup of the simulations was similar to that described in \citet{Kusano_2012}.

\begin{figure}
\includegraphics[width=\linewidth]{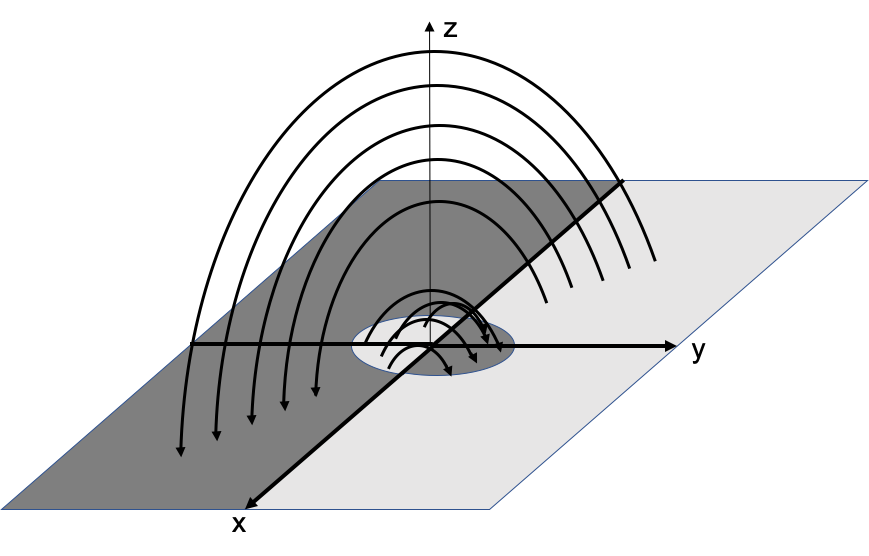}
\caption{Illustration of the simulation setup as in \citet{Kusano_2012}. The solid curved lines with arrows show the magnetic field of both the emerging bipolar field and the overlying force-free field. The white and grey areas on the lower surface indicate positive and negative polarity regions on the photosphere.}
\label{fig:illustration}
\end{figure}

In Kusano et al. the simulations performed have an initial force-free field that extends across a central PIL and is sheared from its potential field configuration by an angle $\theta_0$. A smaller bipole field is then injected from beneath the photosphere and across the PIL representing an ascending magnetic torus. The injected field is rotated at an azimuthal angle of $\psi_e$ relative to a field with its toroidal axis along the PIL. Depending on the parameters $\theta_0$ and $\psi_e$ the ascension of the injected field triggers a flare. An illustration of this setup showing the force-free field and injected BR is shown in \cref{fig:illustration}.

In our study we fix the shear angle of the overlying field to $\theta_0 = 80^\circ$ and the orientation of the emerging field to $\psi_e = 180^\circ$. This scenario produces a powerful flare in \citet{Kusano_2012}. We want to determine whether the motion of the injected field will affect the strength of this flare.

For each simulation the domain size used is $(-2,-0.5,0)\leq(x,y,z)\leq(2,0.5,2))$, this represents a region of the solar atmosphere above the photosphere. The altitude is represented by $z$ and the x-axis $(y=z=0)$ corresponds to the PIL across which the vertical component of the magnetic field $B_z$ changes sign.

The initial field is a force-free field that extends across a central PIL and is sheared from its potential field configuration by an angle $\theta_0$. This is prescribed by,

\begin{equation}
\mathbf{B_{init}} = \exp(-Kz)
        \begin{pmatrix}
		\displaystyle{(\alpha/k)\cos(ky)} \\
		\displaystyle{-(K/k)\cos(ky)} \\
		\displaystyle{\sin(ky)}
		\end{pmatrix}
\end{equation}
where $k=3\pi$ and $K=\sqrt{k^2-\alpha^2}$. 

This field is defined by the parameter $\alpha$ which is related to the shear angle of the field by $\theta_0 = \tan^{-1}(\alpha/K)$. When $\alpha=0$ the field is a potential field. For our simulations $\theta_0$ is fixed at $80^\circ$. 

After each simulation begins the small bipolar field is quickly injected into the force-free field $B_{\text{init}}$ from the lower boundary by prescribing the emerging field $B_e$ and velocity at the boundary. The ascending field is a sphere with radius $r_e$ with constant field intensity $B_e$ rising from its initial centre $(0,0,-r_e)$. The sphere ascends with a constant velocity $v_e$ for a period between $0\leq t\leq\tau_e$. For the simulations in this study the following values were used in nondimensional units $r_e = 0.2$, $B_e = 2.0$, $v_e = 0.01$, and $\tau_e = 18.0$. This gives a radius for the injected field of 4 Mm and allows the injection to halt just before the centre of the torus emerges. 

As with \citet{Kusano_2012} the model for flux emergence is only a convenient mechanism to inject the emerging flux into the active region. What interests us in this case is the effect of horizontal motion of the already injected flux on the subsequent dynamics. 

Just as in \citet{Kusano_2012} the background resistivity was set to $\eta_0 = 10^{-5}$ with a critical current $J_c = 50$ above which the resistivity is enhanced to $\eta_c = 5\times 10^{-4}$, unlike \citet{Kusano_2012} no background viscosity was used as this is not possible in Lare3d. The grid resolution for these simulations is $400\times100\times200$. The results at this resolution were validated by comparing the values for peak KE to those in \citet{Kusano_2012}, which gave similar values.

The same boundary conditions (BC) were used for all simulations. With the exception of the region of the emerging BR on the lower boundary, the BC for velocities was static at all boundaries i.e. $\mathbf{v} = 0$ and the BC for the magnetic field was reflective at all boundaries. These are the default BCs used by Lare3d when the user selects BC\_USER in the control file.

Each simulation was run for until $t=50\;\tau_\text{A}$, where $t_\text{A}$ is the characteristic Alfv\'en time within the simulation.

For a better comparison with \citet{Kusano_2012} variations in density and pressure were neglected in our simulations. The initial density and pressure were set constant across the domain. To prevent radical increases in the pressure caused by ohmic heating, the density and internal energy were fixed across the domain by resetting their values at each timestep.

\begin{figure}[t!]
  \includegraphics[width=\linewidth]{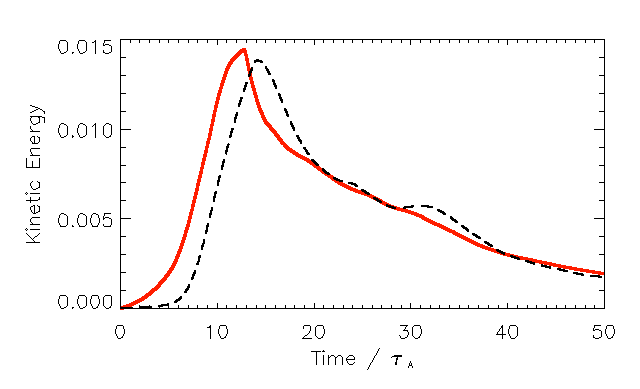}
\caption{Graphs of KE against time for control simulations comparing the setup with density and internal energy fixed throughout the domain (black dashed line) and the compressible simulation with density and internal energy fixed only at the boundary (red solid line).}
\label{fig:ke_comp}
\end{figure}

A control simulation using these conditions and a fully compressible simulation, in which internal energy and density were only fixed at the boundary, were performed to check the effect on the results. The simulations showed  similar flare dynamics, despite dramatic increases in pressure for the compressible simulation. A comparison of these simulations is shown in \cref{fig:ke_comp}, we can see that the compressible simulation peaks slightly earlier but has a similar peak KE.

The core-solver in Lare3d performs calculations in nondimensional units with a normalization defined by $B_0,\;L_0$ and $\rho_0$  \citet{lare}. The normalization used in \citet{Kusano_2012} was $B_0 \sim 0.05$ T, $L_0 \sim 20$ Mm, and $\rho_0 \sim 0.8 \times 10^{-12}$ kg m\textsuperscript{-3} giving a velocity and time normalization of $V_\text{A} \sim 50$ Mm s\textsuperscript{-1}, $\tau_\text{A} \sim 0.4$ s.

Using this normalization, however, gives unrealistically fast speeds for the motion of the BR at the lower boundary of our simulation. The reason for this is that the photospheric density is much higher than the coronal density being about $\rho \sim 3 \times 10^{-4}$ kg m\textsuperscript{-3}. For this reason we instead use the normalization $B_0 \sim 0.05$ T, $L_0 \sim 20$ Mm, and $\rho_0 \sim 3 \times 10^{-4}$ kg m\textsuperscript{-3}  giving $V_A \sim 2.5$ Km s\textsuperscript{-1} and $\tau_A \sim 7766.5 s \approx 130$ minutes.

The chosen normalization gives the correct density at the lower boundary of our domain and therefore
realistic speeds for the motion of our BR. The density is now, however, far too high in the coronal part of our domain so that the time scales for flare dynamics are much larger than would be expected.

Unfortunately, no normalization will give realistic values across the domain. In order to properly resolve this inconsistency a stratified rather than constant density profile should be used in future studies.

\section{Results}

\subsection{Effect of Linear oscillations}

The initial study considered the effect of linear oscillations of the injected field on the dynamics of reconnection and flare triggering. Linear oscillations both along and across the PIL were considered. The centre of the injected field was driven in the $x$ and $y$ directions as follows:
\begin{align}
  D_x = & A_x\sin(2\pi f t) \\
  D_y = & A_y\sin(2\pi f t)
\end{align}

where $A_x$ and $A_y$ are the amplitude of the driving in the $x$ and $y$ directions expressed in nondimensional length units and $f$ is the frequency of the driving expressed in inverse Alfv\'en times.

For the first set of runs $A_y$ was fixed at the value of $4$ Mm, whilst $A_x$ and the frequency $f$ were varied. The values of $A_x$ used were $0,4,8,16,20$ Mm and the frequencies used were $0.1,0.5,1.0,2.0,5.0\; \tau_A^{-1}$. A control simulation with a non-oscillating injected field was also performed.

\begin{figure}[t!]
  \includegraphics[width=\linewidth]{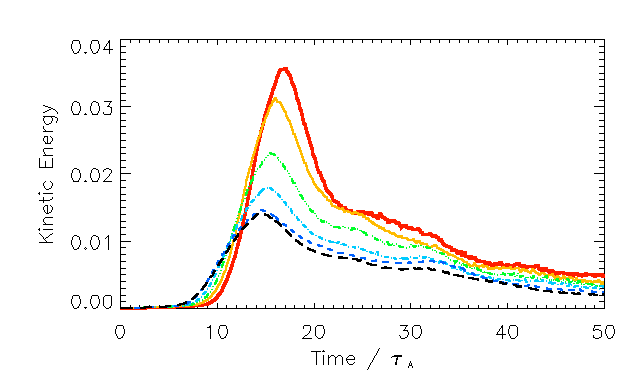}
\caption{Graphs of KE against time for the first set of simulations with $A_y = 4$ Mm and $f = 1 \tau_A^{-1}$. The amplitude $A_x$ is different for each simulation: 0 Mm (black, long dashed line), 4 Mm (dark blue, short dashed line), 8 Mm (light blue, dotted-dashed line) 12 Mm (green, triple dotted-dashed line), 16 Mm (yellow, thin solid line), and 20 Mm (red, thick solid line).}
\label{fig:ke_ax}
\end{figure}

Graphs of kinetic and magnetic energy for these simulations show an increase in KE and a corresponding decrease in magnetic energy as reconnection occurs and a flare is triggered. KE graphs for each simulation at $f = 1 \tau_A^{-1}$ are shown in \cref{fig:ke_ax}. We can see that as that as $A_x$ increases, so does the peak KE.

\begin{figure*}[t!]
\includegraphics[width=\linewidth]{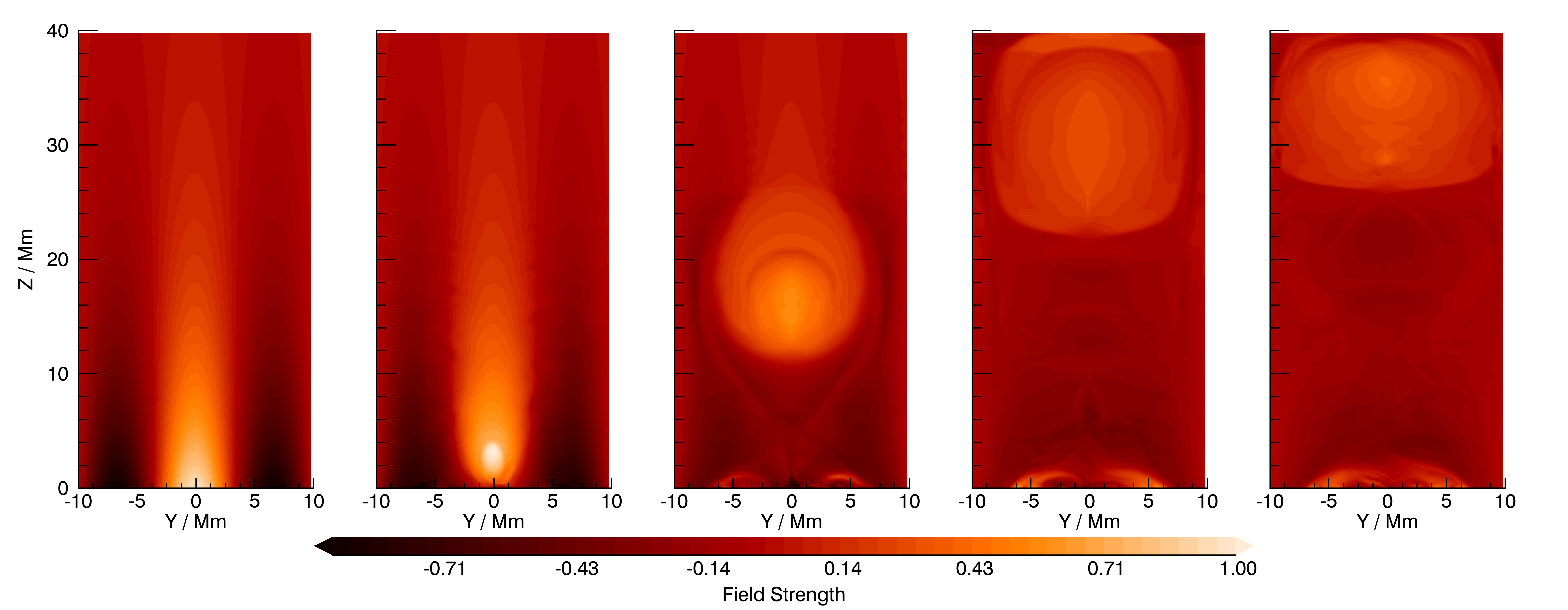}
\caption{Contour plots of $B_x$ through the plane $x=0$ at times 0, 5, 10, 15, and 20 $\tau_A$ for the simulation in which $A_y = 4$, $A_x = 4$ Mm, and $f=1.0\; \tau_A^{-1}$  clearly showing the magnetic reconnection and subsequent eruption.}
\label{fig:recon}
\end{figure*}

Every simulation resulted in an eruptive flare being triggered. \cref{fig:recon} shows the evolution of a flare being triggered for the case $A_x = 4$ Mm, $f=1.0\; \tau_A^{-1}$. 

\begin{figure}
  \includegraphics[width=\linewidth]{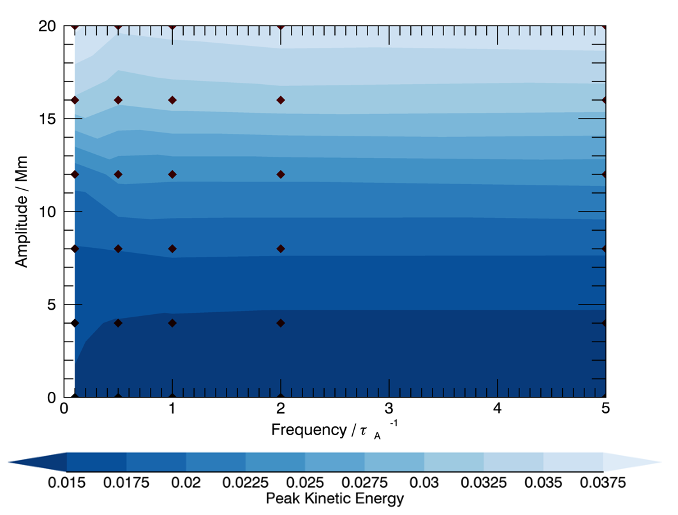}
\caption{Contour plot of the peak KE for different values of amplitude $A_x$ and frequency $f$ with amplitude $A_y$ fixed at $4$ Mm. Each black diamond on the plot is a data point from a simulation. }
\label{fig:ke_a}
\end{figure}

For each simulation the peak KE was recorded. The values for peak KE were similar to those in \citet{Kusano_2012}. For a non-oscillating injected field the peak KE was 0.0138, for all simulation with linearly oscillating injected fields the peak KE was higher. A contour plot showing the peak KE for each simulation is shown in \cref{fig:ke_a}.

The results indicate that the peak KE of the flare is strongly affected by the amplitude of the oscillation but is relatively unaffected by the frequency of the oscillation.

\begin{figure}
  \includegraphics[width=\linewidth]{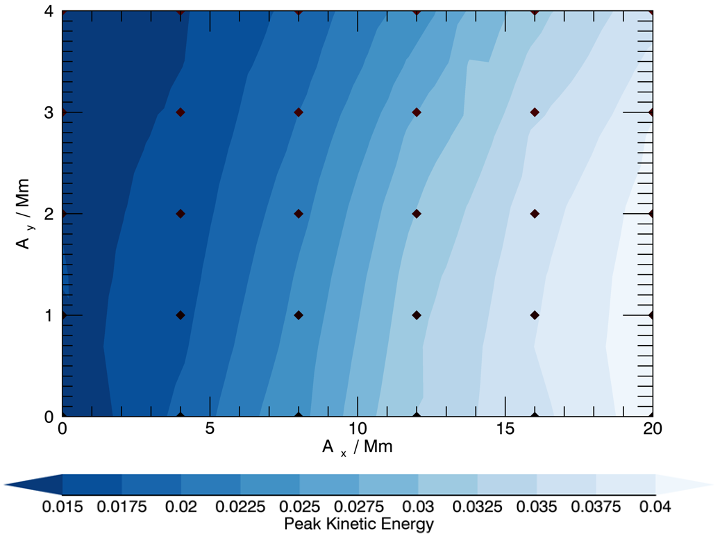}
\caption{Contour plot of the peak KE for different values of amplitudes $A_x$ and $A_y$ with frequency fixed at $f = 1 \tau^{-1}_A$ . Each black diamond on the plot is a data point from a simulation.  }
\label{fig:ke_b}
\end{figure}

Having observed that the peak KE depends on the amplitude $A_x$ of the oscillation along the PIL, the question naturally arose of whether there is a similar dependence on the amplitude $A_y$ of the oscillation across the PIL, or indeed on the direction of the oscillation. To answer this question a second set of simulations was performed. For these simulations the frequency $f$ was fixed at one oscillation per Alfv\'en time and both $A_x$ and $A_y$ were varied. The values of $A_x$ used were again $0,4,8,16,20$ Mm, the values of $A_y$ used were $0,1,2,3,4$ Mm. The amplitudes $A_y$ are smaller so that the BR remains within the domain, which is shorter in the $y$-direction.  A contour plot showing the peak KE for each simulation is shown in \cref{fig:ke_b}.

The results indicate that whilst oscillation along the PIL can substantially increase the peak KE released in a flare,  oscillation across the PIL actually detracts from the peak KE released. This makes sense as when the injected field is not on the PIL it will no longer oppose the surrounding field at the the photosphere at all points.

\subsection{Effect of Torsional oscillations}

The effect of torsional oscillations of the injected field of the dynamics of flare triggering was also studied in a set of simulations. The torsional oscillations were driven by altering the orientation of the emerging field $\psi_e$ as follows:
\begin{equation}
  \psi_e = 180^\circ + A_\psi \sin(2\pi ft)
\end{equation}

The amplitude of oscillations $A_\psi$ as well as the frequency $f$ were varied. The values used for $A_\psi$ were $30^\circ, 60^\circ, 90^\circ, 120^\circ, 150^\circ, 180^\circ$, the frequencies used were $0.1,0.5,1.0,2.0,5.0\; \tau_A^{-1}$. 

The simulations were extensive with $\theta_0$ also being varied for some of the runs. It was found that neither the frequency nor amplitude of torsional oscillations had any major effect on the peak KE.

 \begin{figure}[t]
  \includegraphics[width=\linewidth]{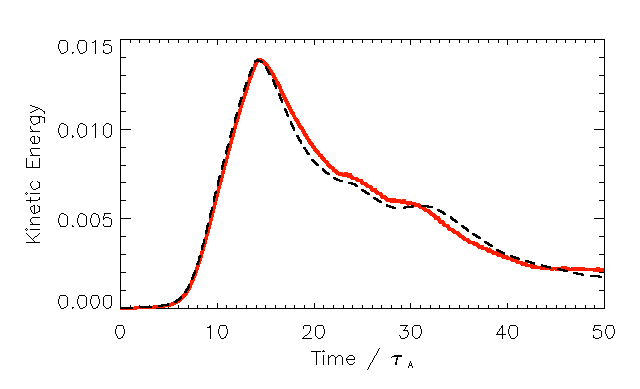}
\caption{Graphs of KE against time for a non-oscillating injected field (black, dashed line) and a torsionally oscillating injected field  with an amplitude $A_\psi = 90^\circ$ and frequency $f = 1\; \tau_A$ (red, solid line).}
\label{fig:ke_torsion}
\end{figure}

\cref{fig:ke_torsion} illustrated a typical example of the results from these simulations. In this particular simulation $A_\psi = 90^\circ$ and $f=1.0\; \tau_A^{-1}$. These graphs show that there is very little difference in the dynamical evolution between the scenario with a non-oscillating injected field and the scenario with a torsionally oscillating injected field. The peak KE for the scenario with a non-oscillating injected field was 0.0138 whereas the peak KE for the scenario with a torsionally oscillating injected field was 0.0139. Similar results were found for all other simulations with torsional oscillations.

\subsection{Effect of Collisions}

\begin{figure}
  \includegraphics[width=\linewidth]{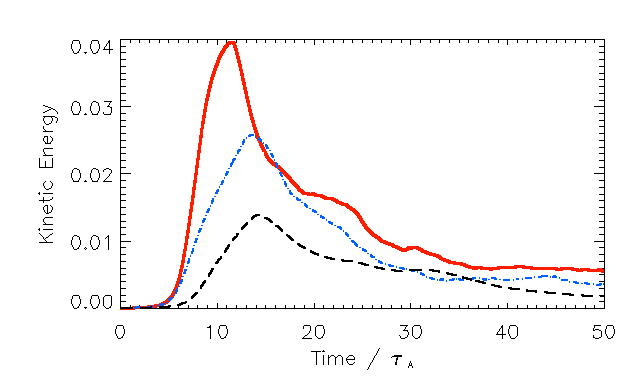}
\caption{Graphs of KE against time for a control simulation with a single emerging BR (black, dashed line), a simulation with two noncolliding BRs (blue, dotted-dashed line) and a simulation of two colliding BRs (red, solid line).}
\label{fig:ke_collide}
\end{figure}

The final set of simulations performed each had two injected fields instead of one. Both of these fields still had an orientation of $\psi_e = 180^\circ$ and the same parameters as the rest of the study. Graphs of total KE against time are shown for both these simulations and for the control simulation with a single BR in \cref{fig:ke_collide}.

\begin{figure*}[htb!]
  \begin{center}
 \minipage{\textwidth}
   \centering
   \includegraphics[width=0.46\linewidth]{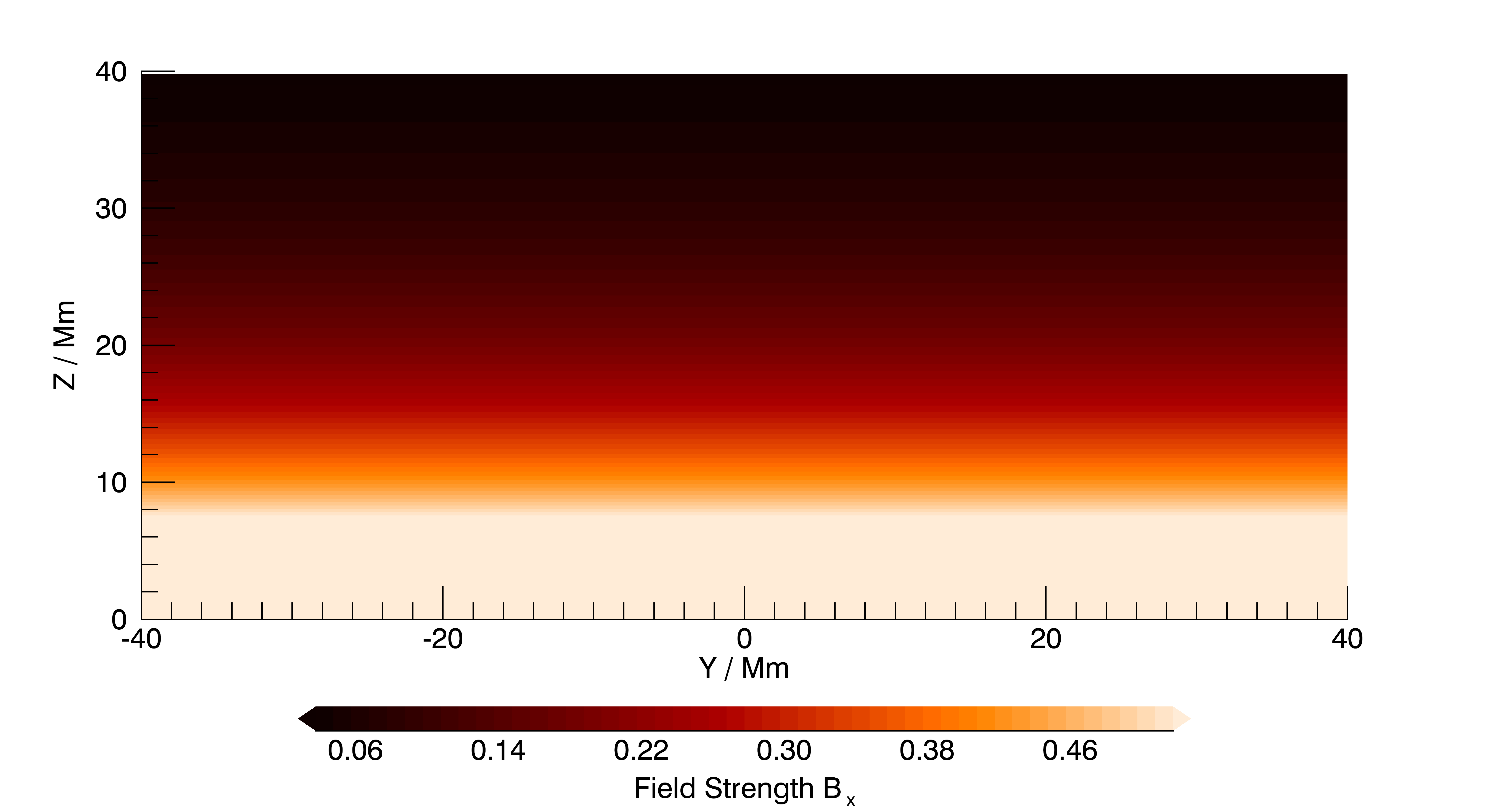}
   \hspace{0.05\linewidth}
   \includegraphics[width=0.46\linewidth]{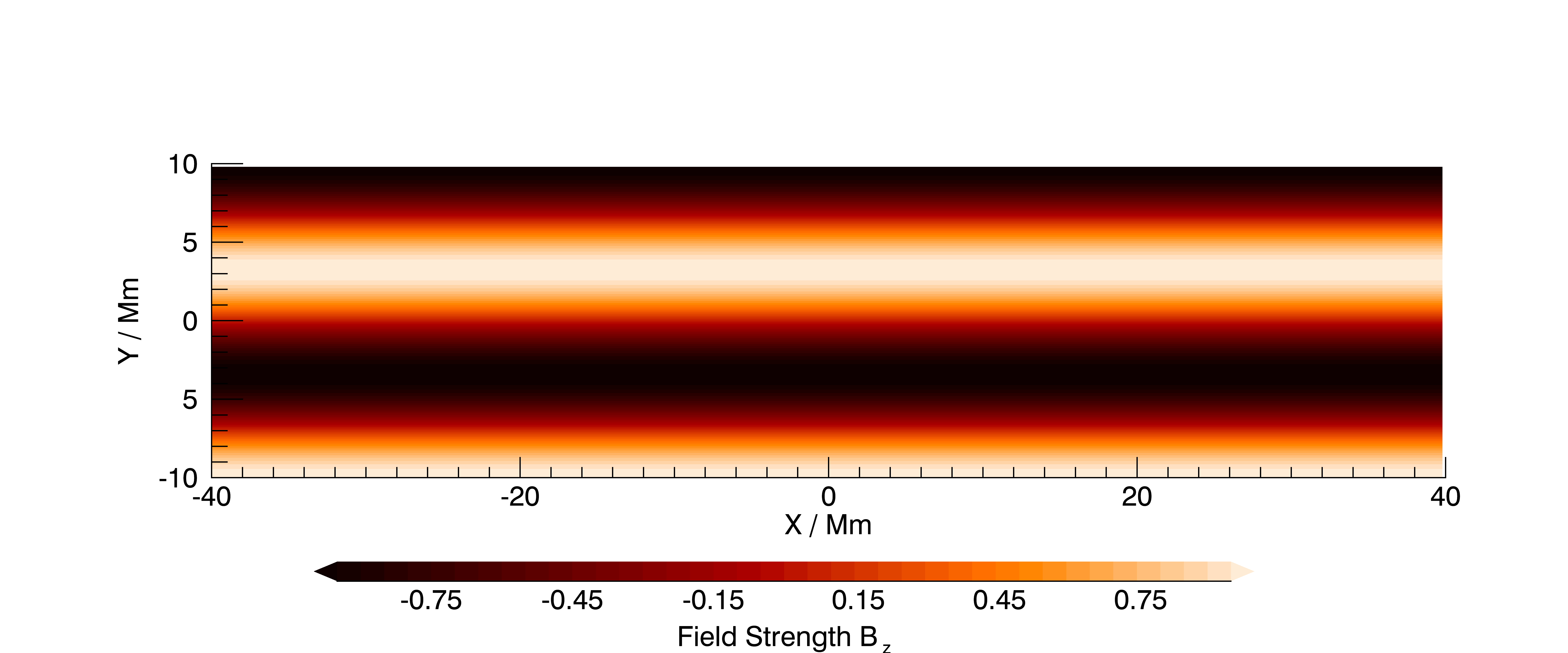}
 \endminipage\hfill
 \minipage{\textwidth}
 \centering
   \includegraphics[width=0.46\linewidth]{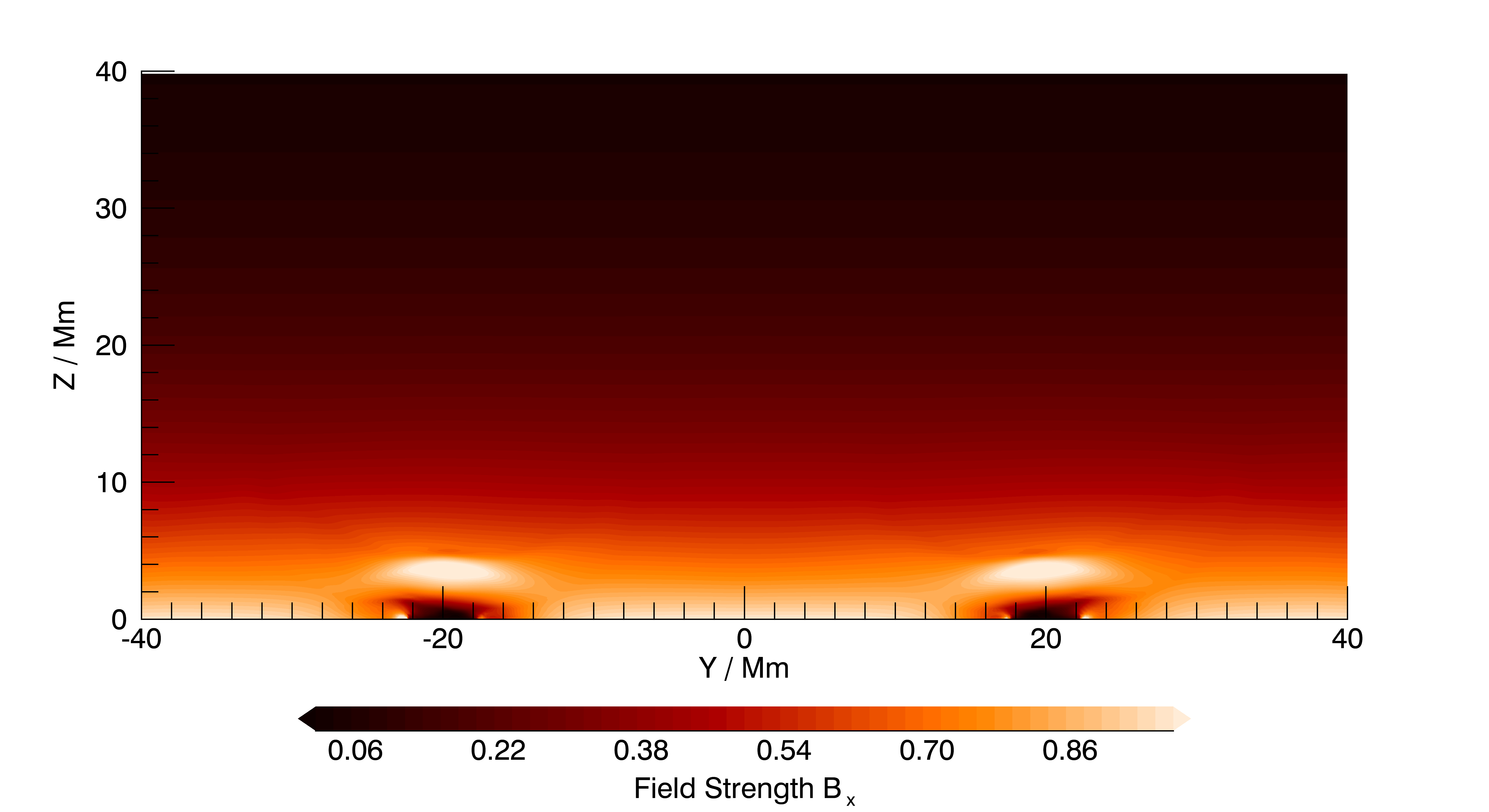}
   \hspace{0.05\linewidth}
   \includegraphics[width=0.46\linewidth]{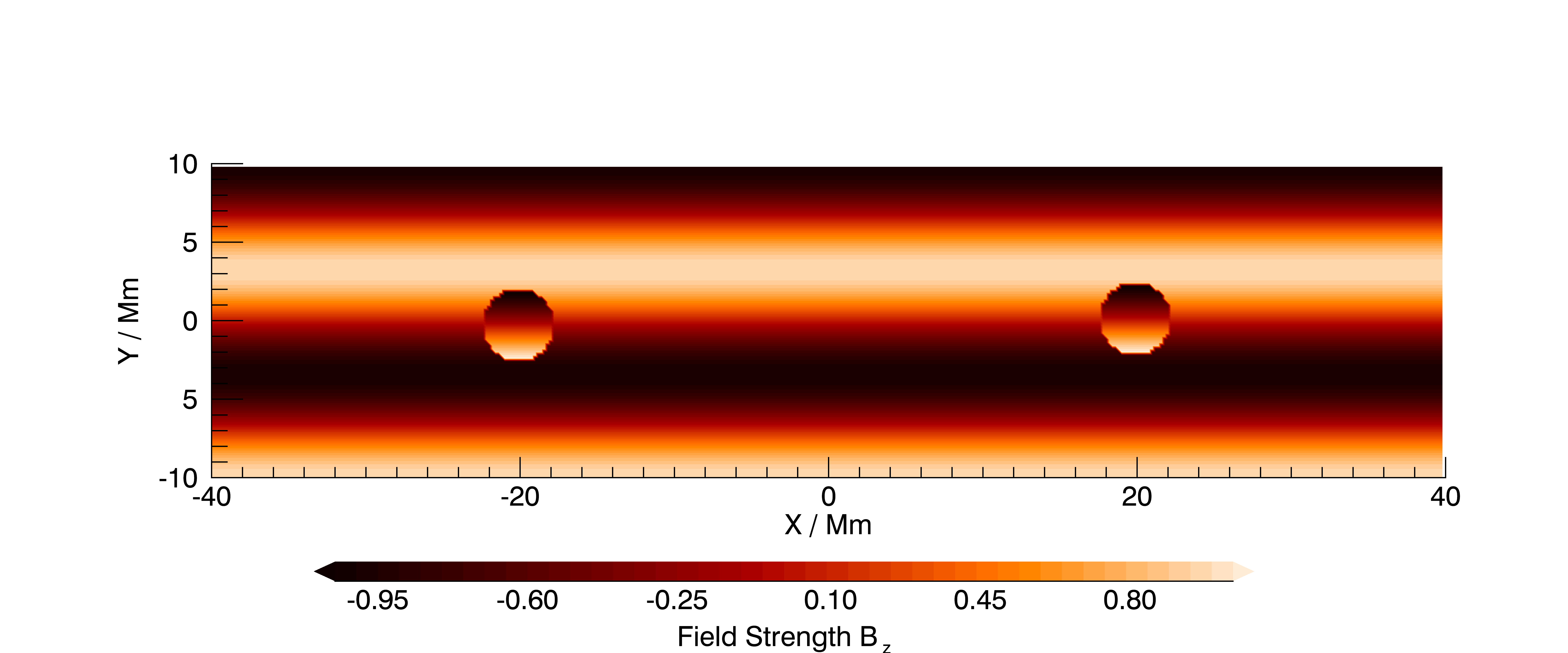}
 \endminipage\hfill
 \minipage{\textwidth}
 \centering
   \includegraphics[width=0.46\linewidth]{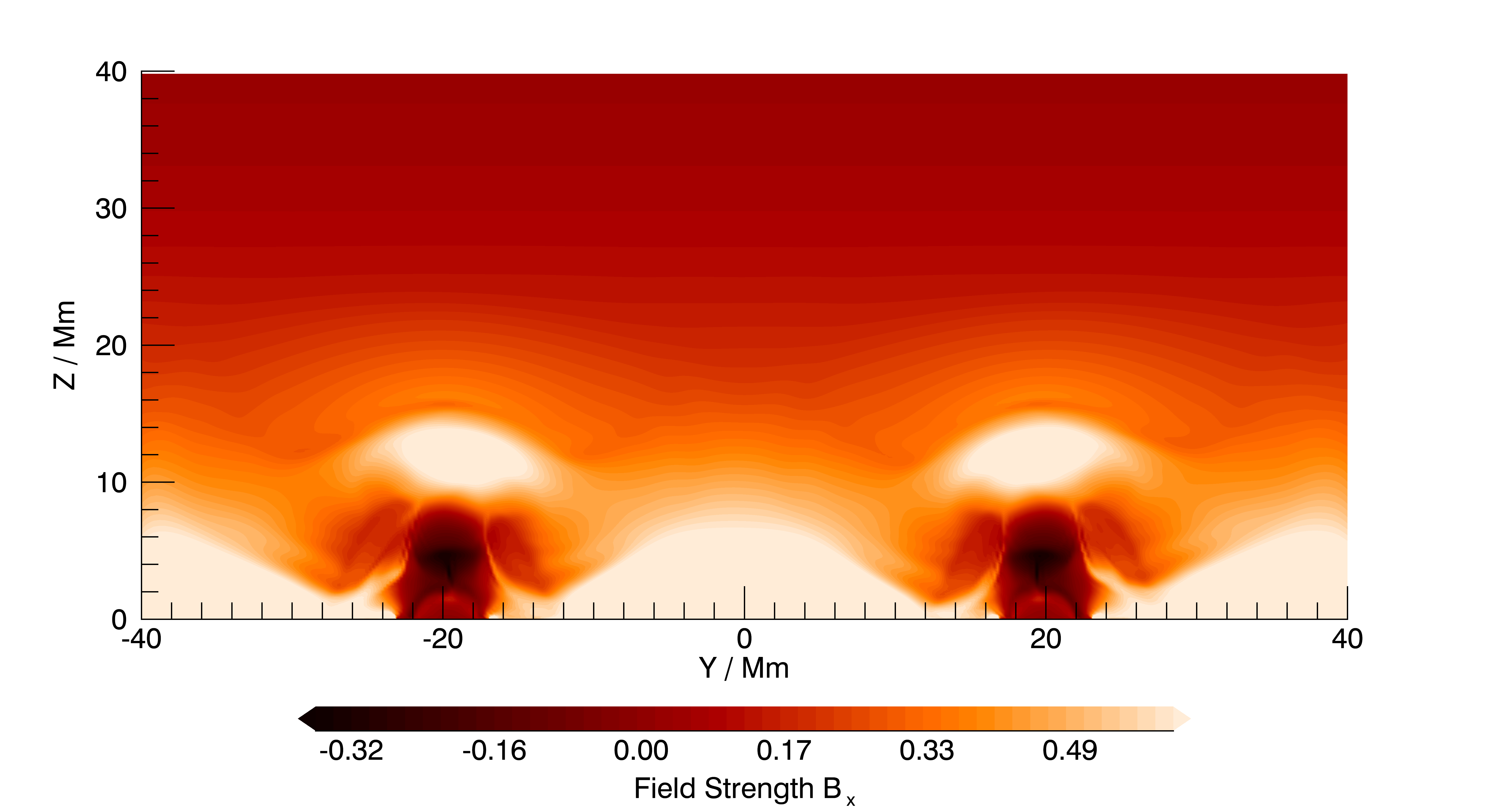}
   \hspace{0.05\linewidth}
   \includegraphics[width=0.46\linewidth]{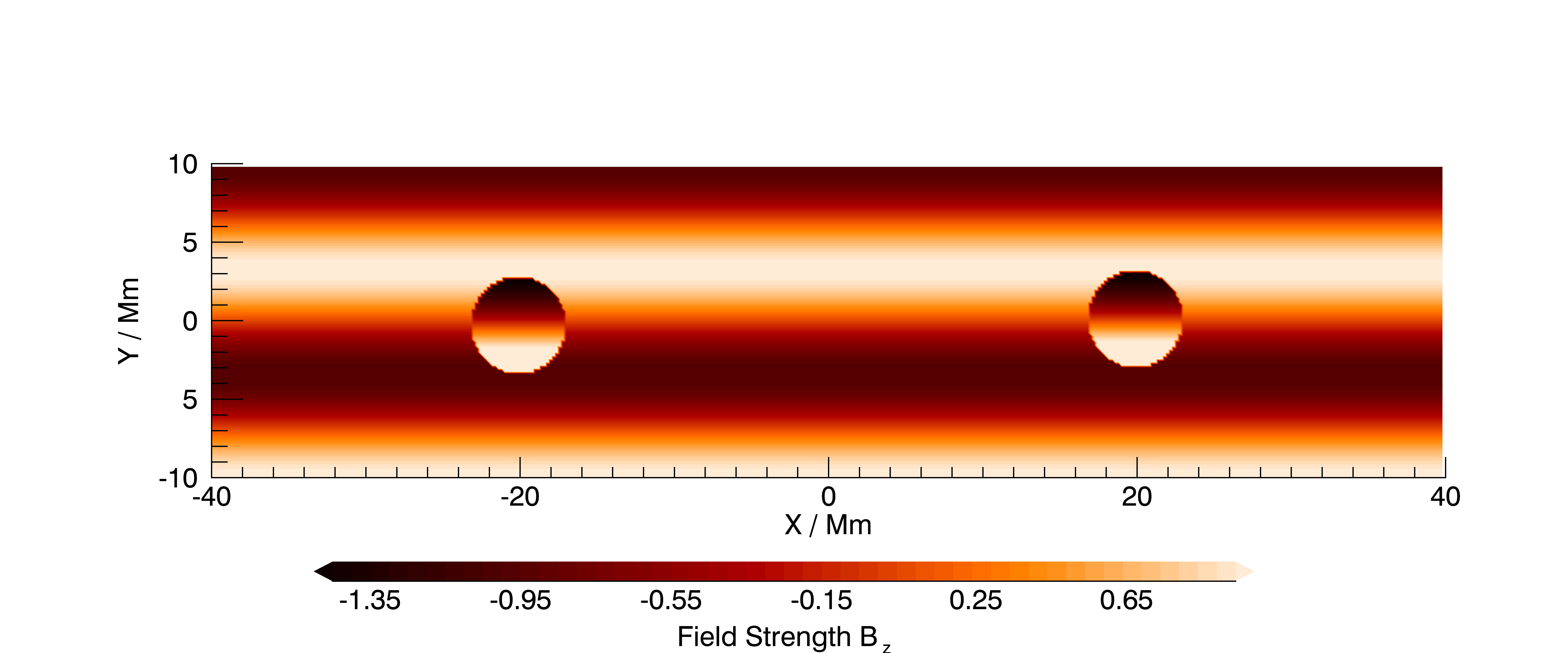}
 \endminipage\hfill
 \minipage{\textwidth}
 \centering
   \includegraphics[width=0.46\linewidth]{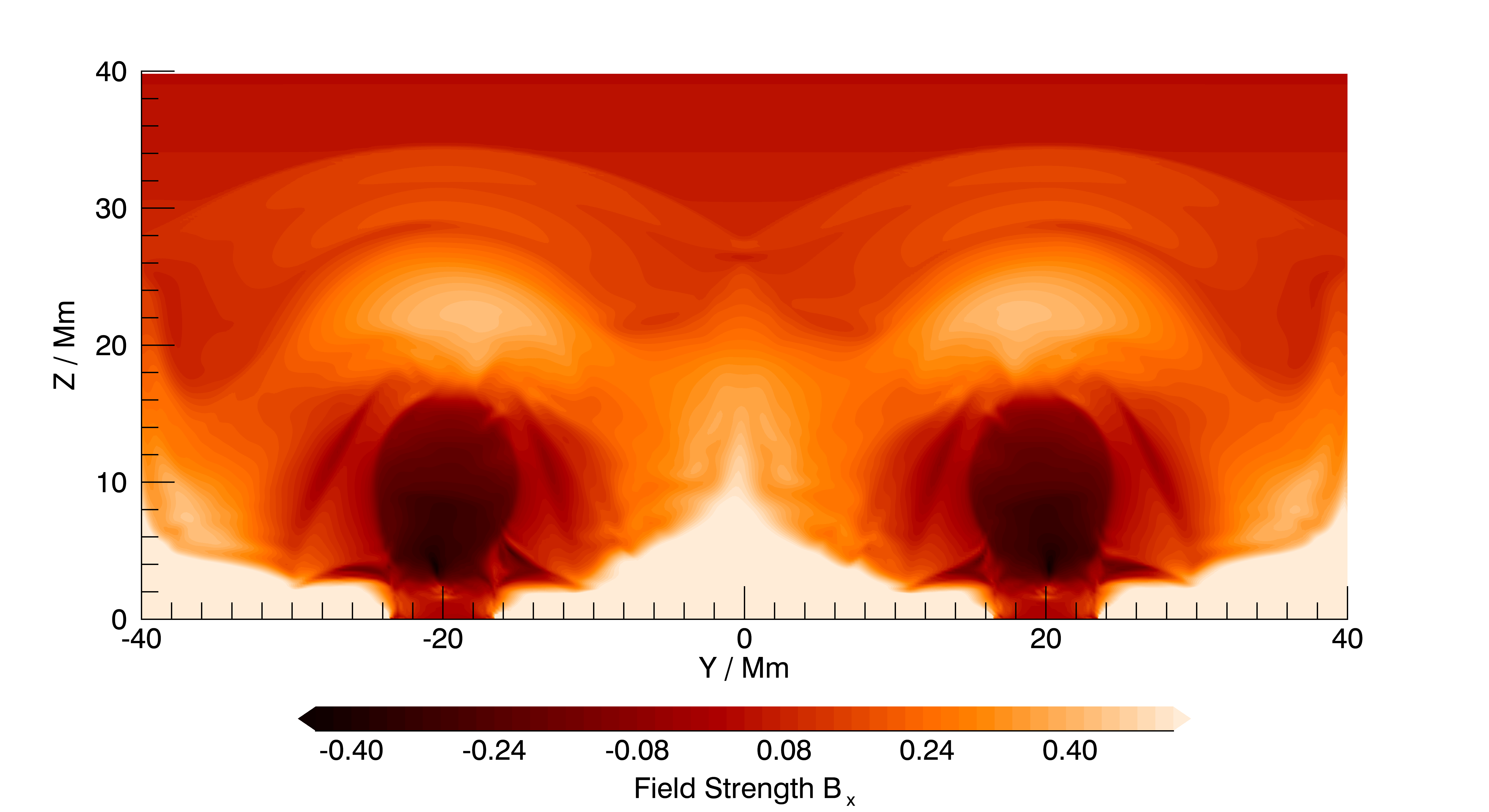}
   \hspace{0.05\linewidth}
   \includegraphics[width=0.46\linewidth]{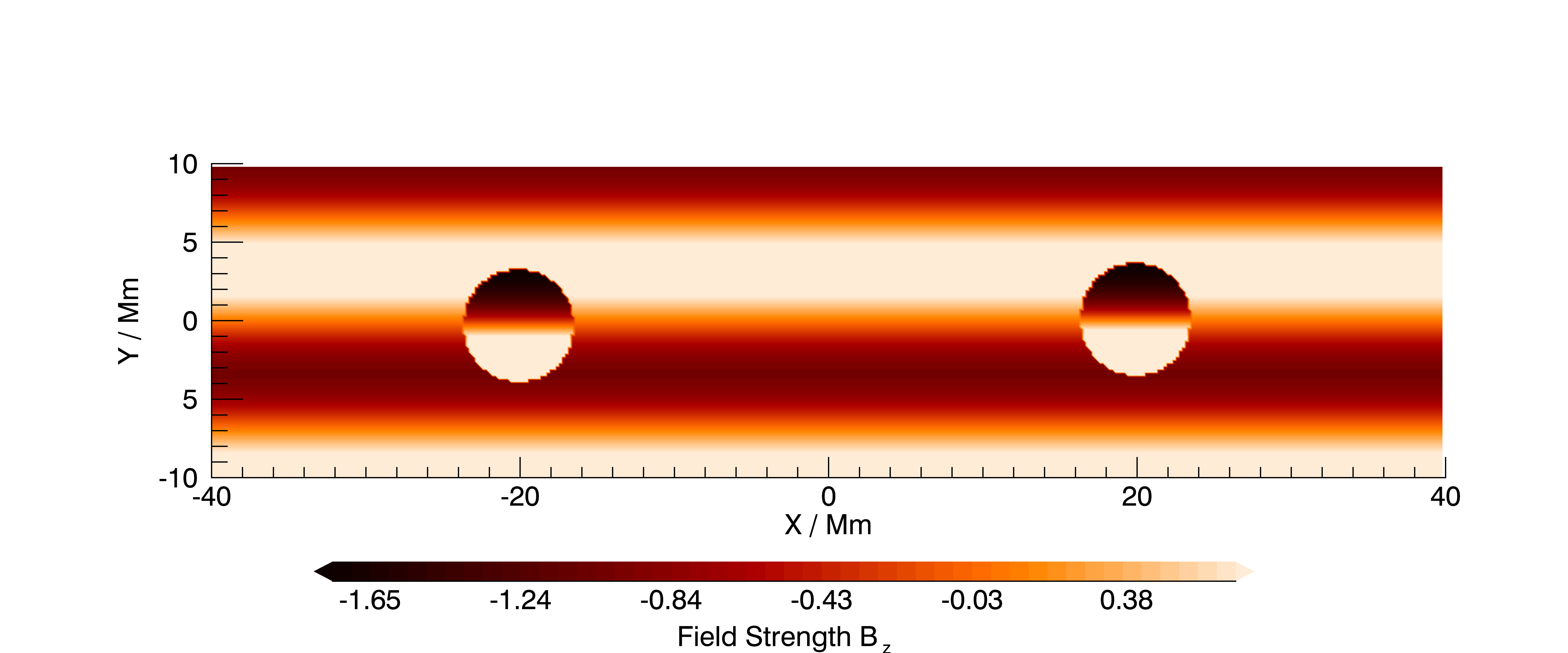}
 \endminipage\hfill
 \minipage{\textwidth}
 \centering
   \includegraphics[width=0.46\linewidth]{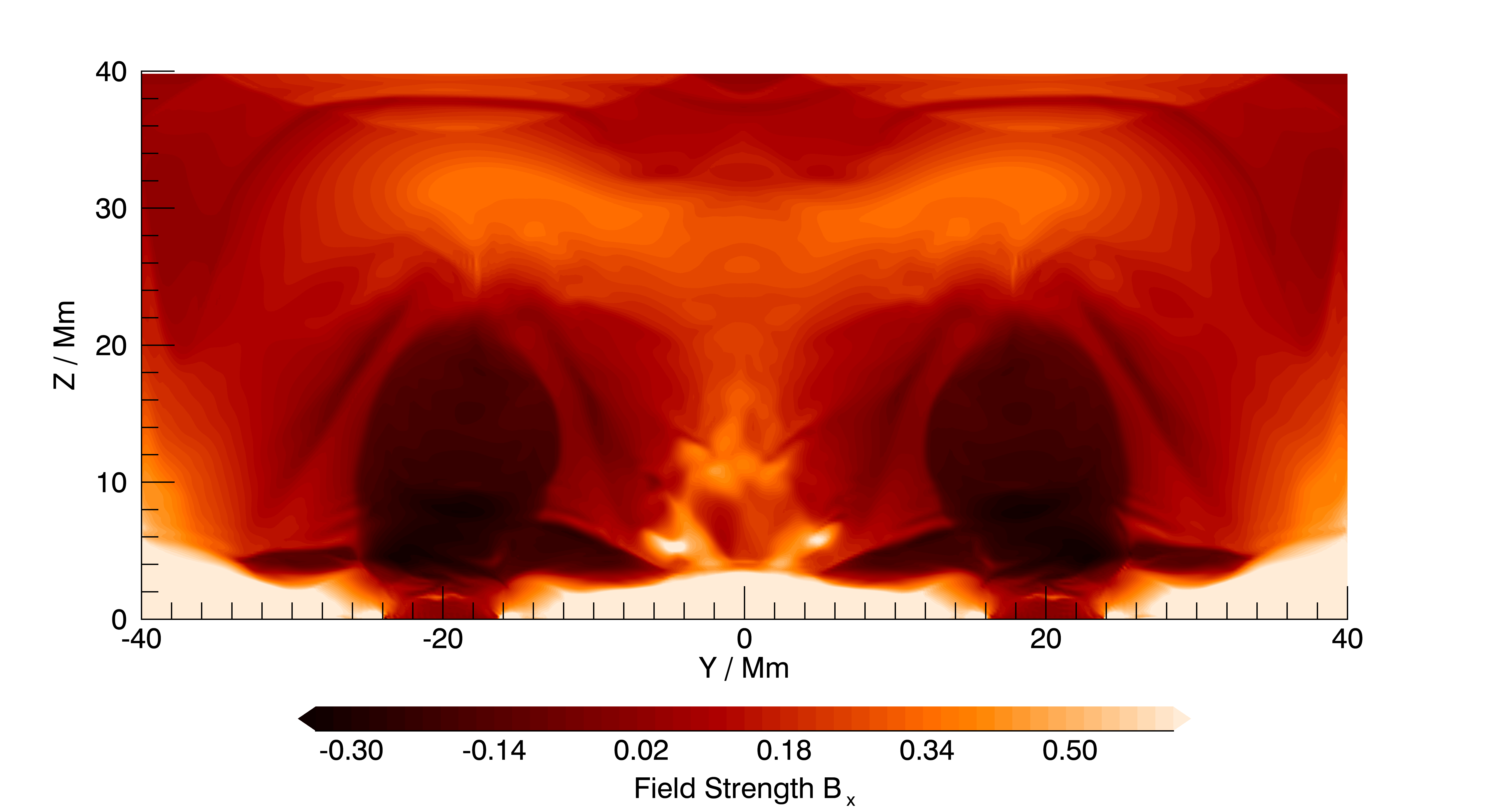}
   \hspace{0.05\linewidth}
   \includegraphics[width=0.46\linewidth]{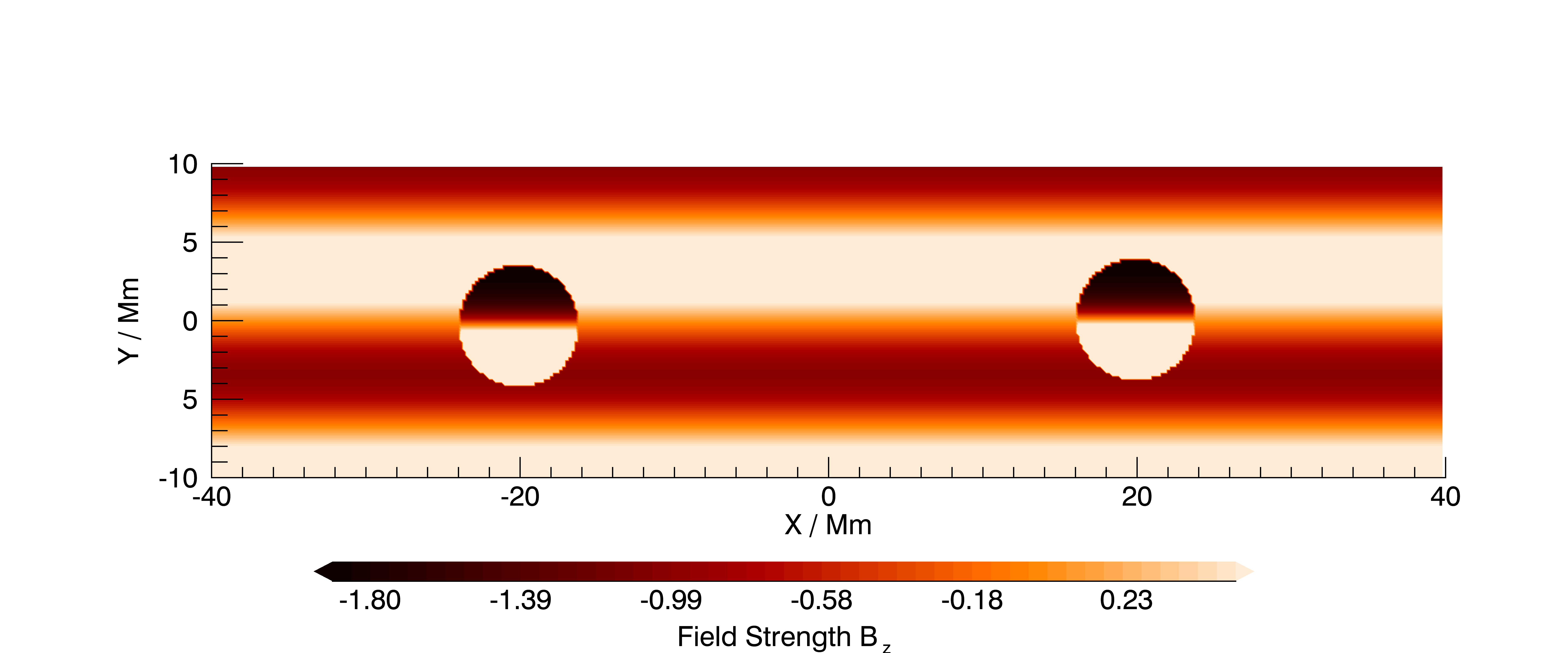}
 \endminipage\hfill
 \end{center}
 \caption{Contour plots of (left) $B_x$ through the plane $y=0$ and (right) $B_z$ through the plane $z=0$ at times 0, 3.6, 7.2, 10.8, and 14.4 $\tau_A$ for the simulation with two noncolliding small-scale BRs. The magnetic reconnection and subsequent eruptions can be seen occurring at the location of each injected field.}
 \label{fig:recon_double}
 \end{figure*}

In the first simulation the injected small-scale BRs were placed $40$ Mm apart along the PIL and each displaced $0.2$ Mm from the PIL in opposite directions. The injected fields were allowed to emerge as usual. Both injected fields independently reconnected with the overlying field producing two distinct eruptions as shown in \cref{fig:recon_double}.

In the second simulation the injected small-scale BRs began in the same positions but were given constant velocities of $2$ Mm per $\tau_A$ toward the centre of the domain and each other. During the simulation the injected fields collide. During the collision the fields exist in a linear superposition at the lower boundary but continue to be modelled by resistive MHD within the domain. After colliding the injected fields continue to move along the lower boundary until they exit the domain altogether.

The injected fields begin to reconnect with the overlying field independently as they emerge, more of the field is then reconnected as the small-scale BRs begin to move together, and this culminates in a single large eruption  as shown in \cref{fig:recon_collide}.

The peak KE released in the simulation with two noncolliding, nonrotating BRs is roughly double the peak KE seen in the simulation of a single, non-oscillating BR, which is what we might expect. The peak KE released for the simulation with two colliding, nonrotating BRs is, however, much larger than for two noncolliding BRs showing that the strength of the flare has been enhanced by either the collision or the motion of the two BRs, or a combination of these effects. In either case the result of the collision is that the previously separated flares caused by each of the BR have now coalesced into one much larger flare.

A further four simulations were performed to investigate the effects of rotation of the small-scale BRs during collision. The noncolliding simulation was repeated firstly with the BRs corotating and then counter-rotating. Then the colliding simulation was repeated firstly with the BRs corotating and then counter-rotating. All rotation was at at a constant rate of one revolution per $\tau_A$. The results of these simulations were particularly interesting. The results of all six simulations are shown in \cref{table:1}. \\

\begin{table}[h]
\begin{tabular*}{\columnwidth}{c@{\extracolsep{\fill}}cc}
 \hline
\hline
Movement & Rotation & Peak KE  \\ [0.5ex] 
\hline
Noncolliding & Nonrotating & 0.0269 \\ 
Noncolliding & Corotating & 0.0203 \\ 
Noncolliding & Counter-rotating & 0.01536 \\ 
Colliding & Nonrotating & 0.0396 \\ 
Colliding & Corotating & 0.0012 \\ 
Colliding & Counter-rotating & 0.0011 \\ 
\hline
\end{tabular*}
\caption{A Table Showing the Peak KE for Each of the Six Simulations Involving Two Emerging BR Rather Than One.}
\label{table:1}
\end{table}

The peak KE released for the simulations with two noncolliding rotating BRs is slightly less than for the nonrotating case. The reduced peak KE is caused by the fields deviating from the optimal orientation of $\psi_e= 180^\circ$. Looking at the simulations more closely reveals that both injected fields independently trigger flares but at much later times than in the nonrotating case. Reconnection proceeds whenever the orientation of the BR is close to optimal and the continued disruption of the same section of overlying field eventually triggers a flare. The peak KE is higher for the corotating case because the flares are triggered at the same time unlike the counter-rotating case.

The peak KE released for the simulations with two colliding rotating BRs is almost negligible. Looking at the simulation results more closely reveals that no flares are triggered in either simulation. This is likely because no section of the overlying field is disrupted for long enough to trigger a flare. Each rotating BR is in the optimal orientation only for a short time during rotation and never twice for the same section of overlying field.

\section{Discussion}

The results from these simulations indicate that the motion of the emerging small-scale BRs that trigger flares can have a significant effect on the strength of those flares. In the most extreme cases the effect of motion alone caused over a threefold increase in the peak KE released by the flare. 

The proper motion of magnetic features such as these has been observed on the solar surface, for example, sunspots have been seen to drift at speeds of 0.14 km s\textsuperscript{-1} \citet{Herdiwijaya} and moving magnetic features within sunspots move at $\sim 0.5$ km s\textsuperscript{-1} \citet{Zhang}. Rotational motion of magnetic features has also been observed, the rotational velocity of sunspots has been measured at values of up to $\sim 3.8^\circ$ hr\textsuperscript{-1} during their emergence from the photosphere \citet{Zhu}.

Photospheric motions are known to twist the coronal magnetic field into nonpotential configurations as discussed in \citet{Raman}, \citet{Gerrard}, \citet{Wang} and \citet{Yan1}. These magnetic fields can then be triggered resulting in a flare. In these simulations, however, which include a preexisting nonpotential field, we are only interested in how these motions effect the strength of the flare through the triggering process. 

In our first set of simulations we specifically considered linear oscillation of the emerging BR rather than simply drift; though the results indicate that there may be little distinction. Interestingly the frequency at which the emerged BR oscillates does not seem to effect the strength of the flare. It is thought that this is because the frequency of the oscillation ultimately has no effect on how much of the overlying field is exposed to the disruptive trigger of the emerging small-scale BRs.

The orientation of the oscillation on the other hand has been shown to be very important. Motion along the PIL generally increases the strength of the flare whilst motion across the PIL generally detracts from the strength of the flare. This is likely because the position of the emerging field on the PIL is the most disruptive to the overlying field and more of this field is disrupted if there is motion of the BR along the PIL.

The most explosive flare triggered in our study of linear oscillations had an amplitude of 20 Mm along the PIL and at a frequency of one oscillation per $\tau_A$, this equates to a linear speed of 10 km s\textsuperscript{-1}. Whilst this value is well above those quoted it should be noted that decreasing the frequency and therefore linear speed should not significantly change the results of the simulation. In fact when this was repeated at the lower frequency of $0.1 \tau_A^{-1}$ corresponding to a much more reasonable 1 km s\textsuperscript{-1} the peak KE dropped only slightly from 0.041 to 0.038.

In our simulations of torsional oscillations, most of which used a higher rotational velocity than what is observed, the strength of the triggered flare was unaffected when compared to simulations without torsional oscillation. In our simulations of two rotating BRs the rotation acted to reduce the strength of the flares, this is more likely related with the orientation of the BR rather than the rotation. In these cases the BR being away from the optimal orientation for large periods of time led to a staggered release of magnetic energy through reconnection processes.

The collision of sunspots has been observed to both produce magnetic shear in the field \citet{Gaiz} and to trigger solar flares \citet{Yan2}. In \citet{Yan2} the reconnection resulted from the squeezing of magnetic fields with opposite polarities. In our simulation the two BRs had the same polarity and we have showed that reconnection would have occurred with or without the collision due to the opposite polarity of the BR and the overlying field. The strength of the flare, however, was significantly increased by the collision; this is probably a combination of the effects of the motion of each individual BR and of the coalescence of the reconnected field as shown in \cref{fig:recon_collide}.

Interestingly the KE  graphs all show a similar pattern after the flare. The KE does not linearly decay but rather shows a decay superimposed with significant oscillations reminiscent of the quasi-periodic pulsations discussed in \citet{QPP1} and \citet{QPP2} . Further analysis reveals oscillations of two different periods with the shorter period corresponding to the driving of the BR. The longer period oscillation may be due to oscillatory reconnection, whilst the short period oscillations may be a direct measurement of the BR dynamics or may be flare activity modulated by the BR driving \citet{QPP3,QPP4}. 

\section{Conclusions}

\begin{figure*}[htb!]
 \minipage{\textwidth}
   \centering
   \includegraphics[width=0.46\linewidth]{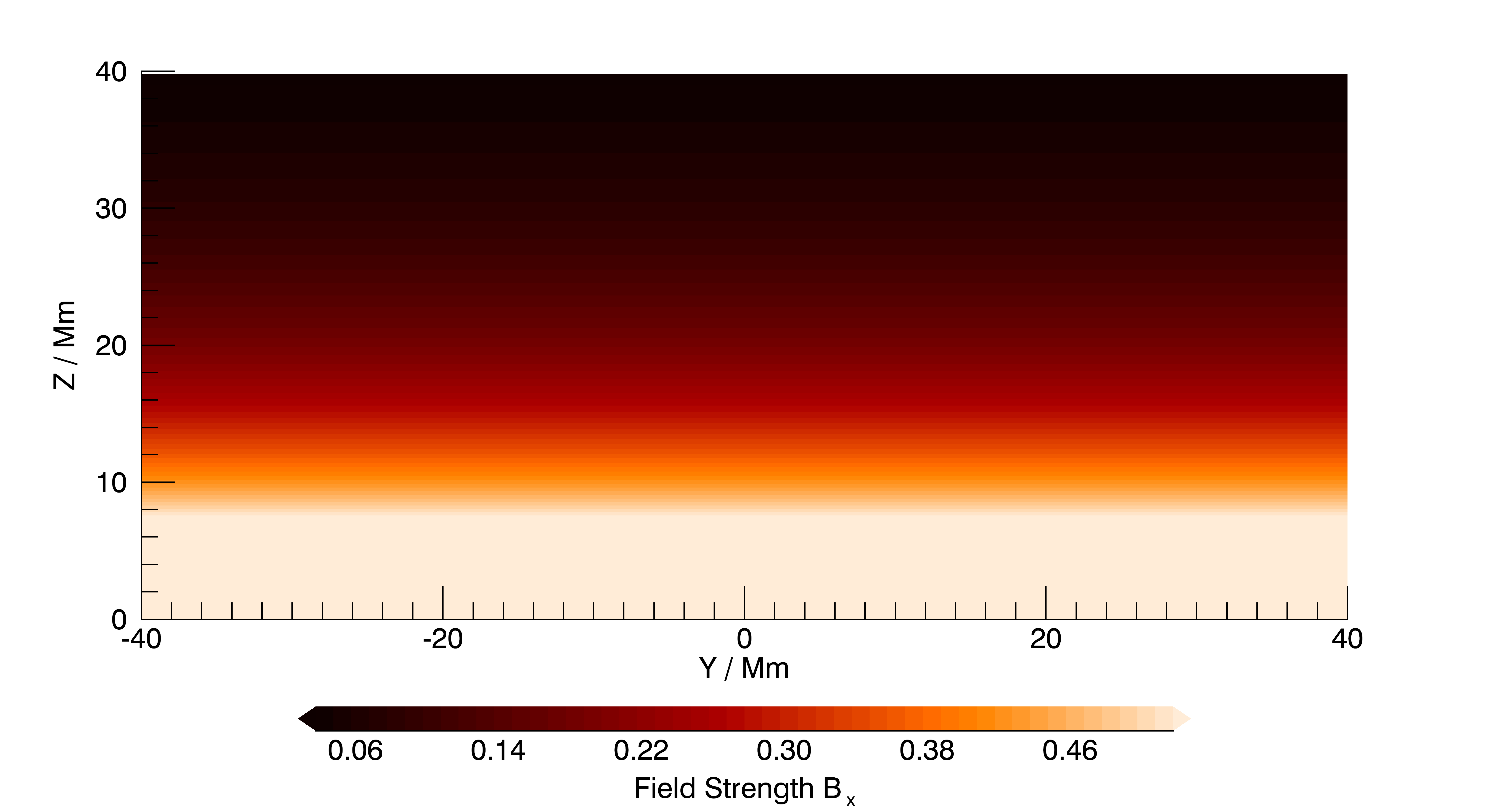}
   \hspace{0.05\linewidth}
   \includegraphics[width=0.46\linewidth]{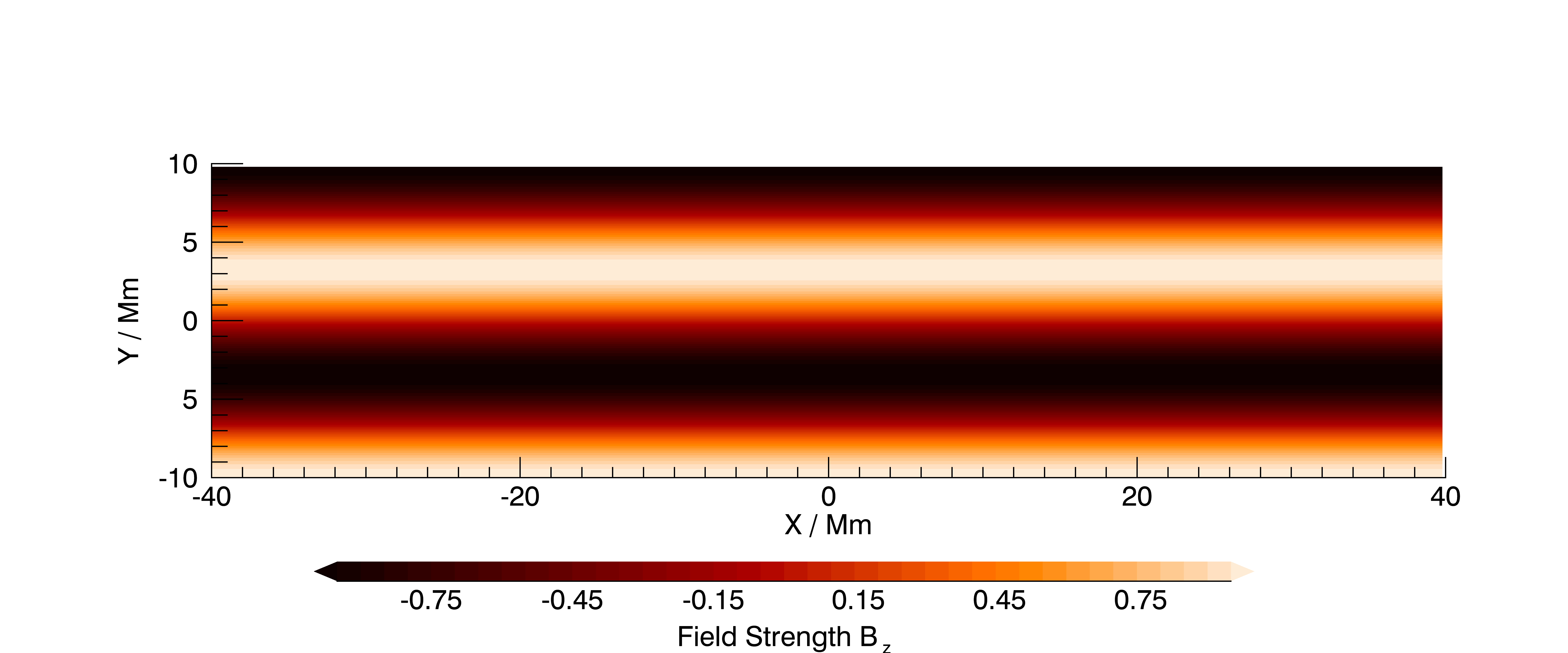}
 \endminipage\hfill
 \minipage{\textwidth}
 \centering
   \includegraphics[width=0.46\linewidth]{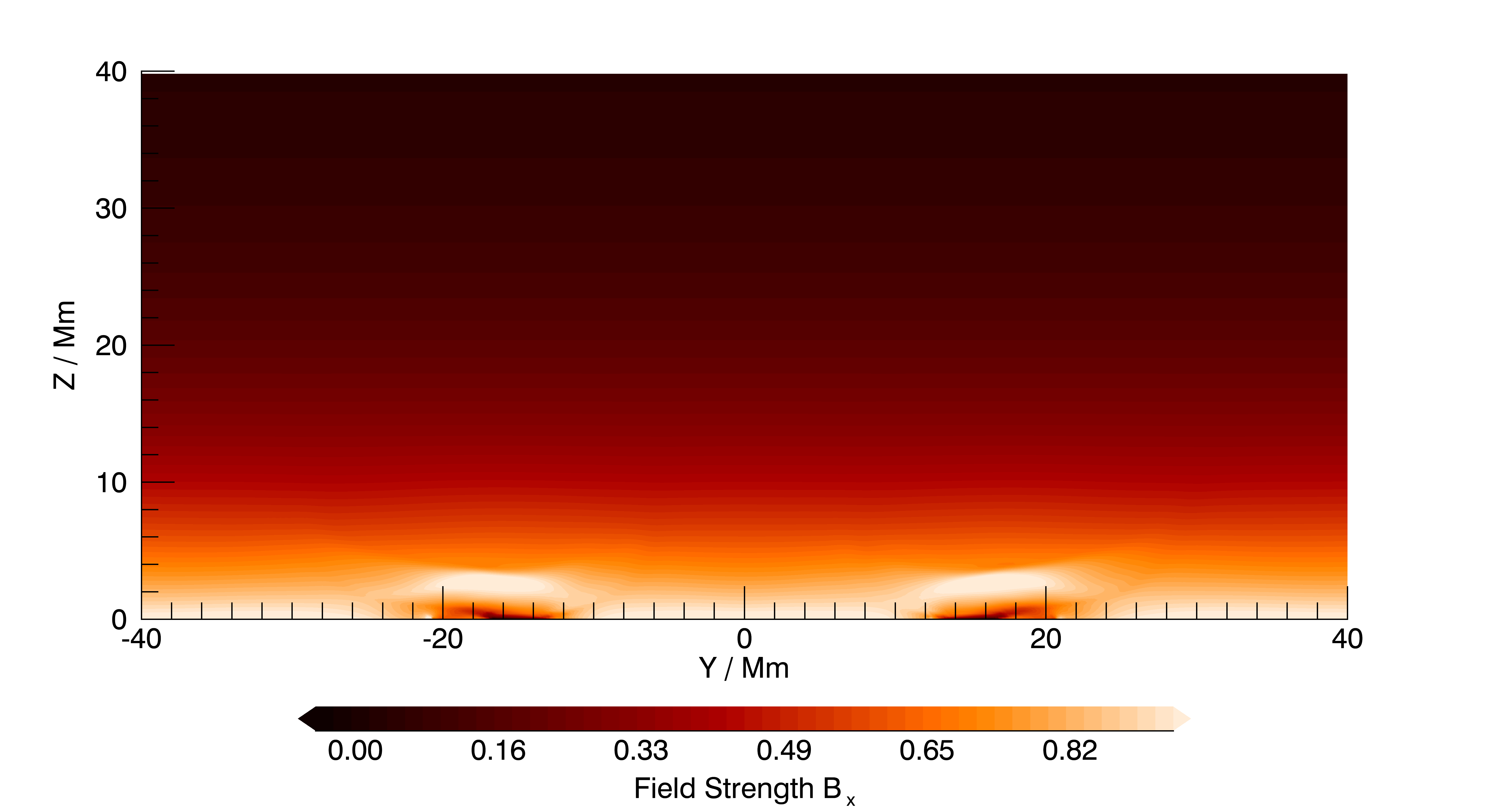}
   \hspace{0.05\linewidth}
   \includegraphics[width=0.46\linewidth]{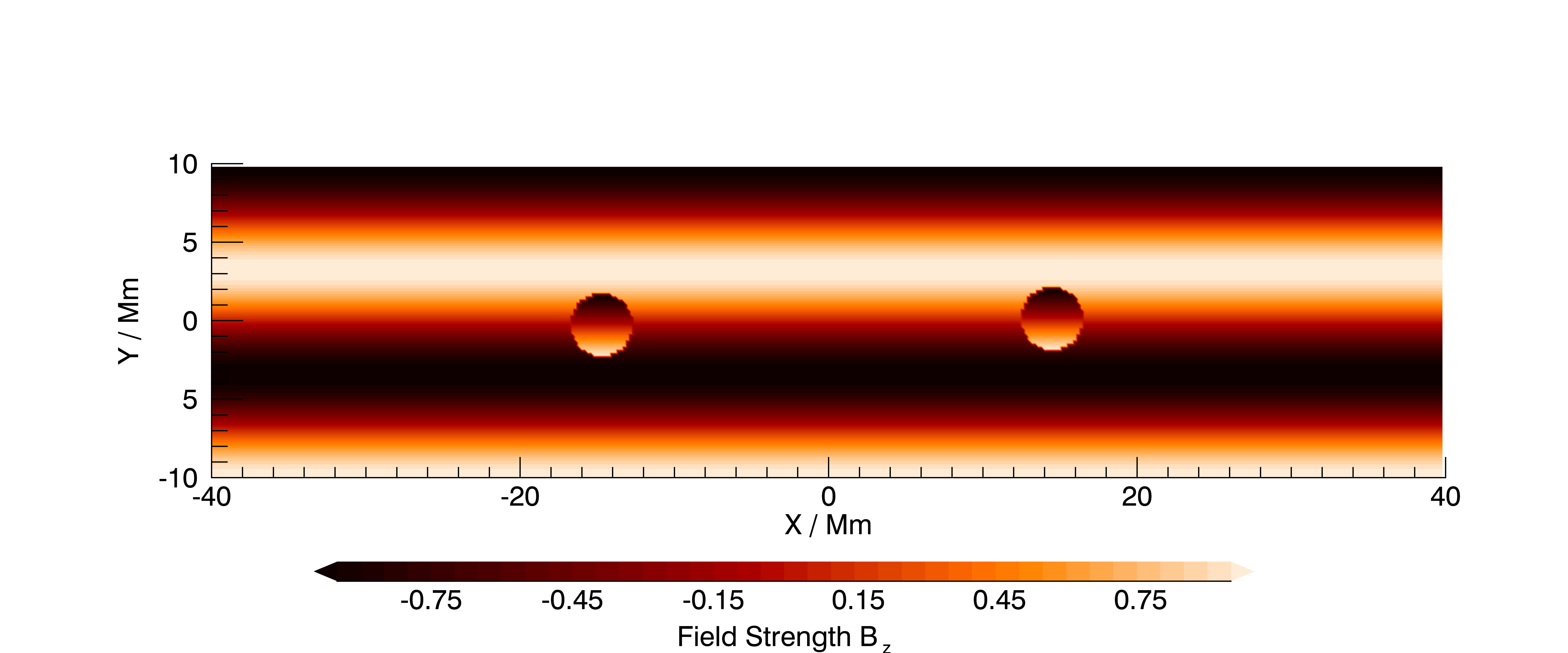}
 \endminipage\hfill
 \minipage{\textwidth}
 \centering
   \includegraphics[width=0.46\linewidth]{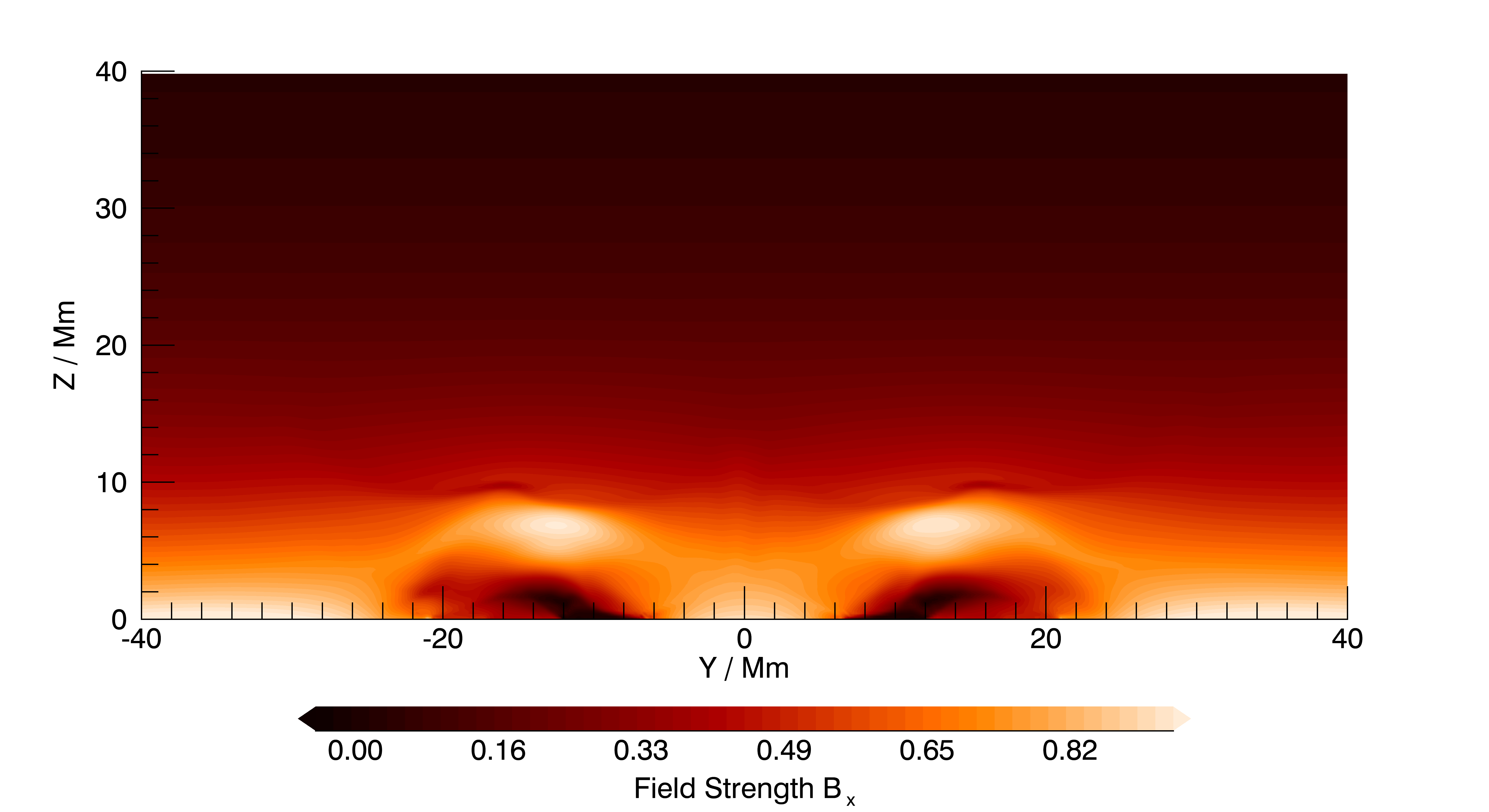}
   \hspace{0.05\linewidth}
   \includegraphics[width=0.46\linewidth]{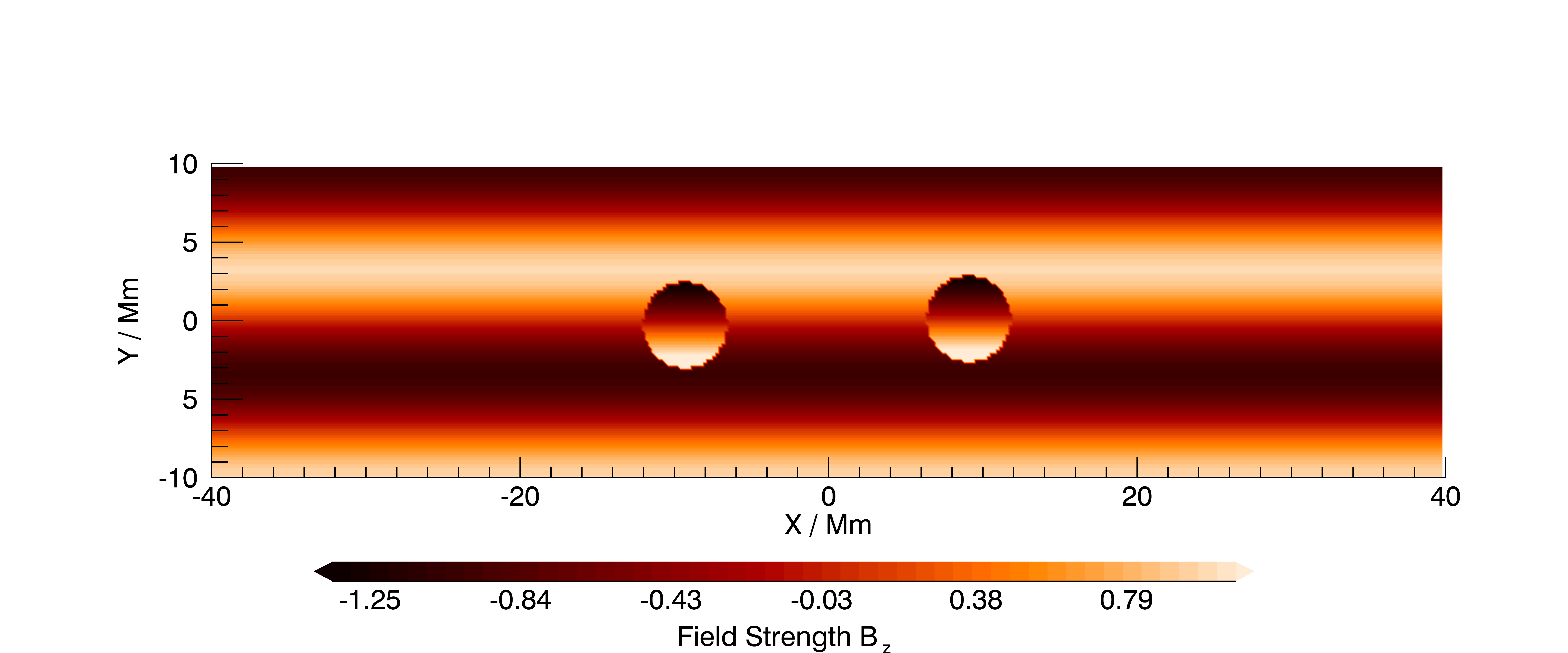}
 \endminipage\hfill
 \minipage{\textwidth}
 \centering
   \includegraphics[width=0.46\linewidth]{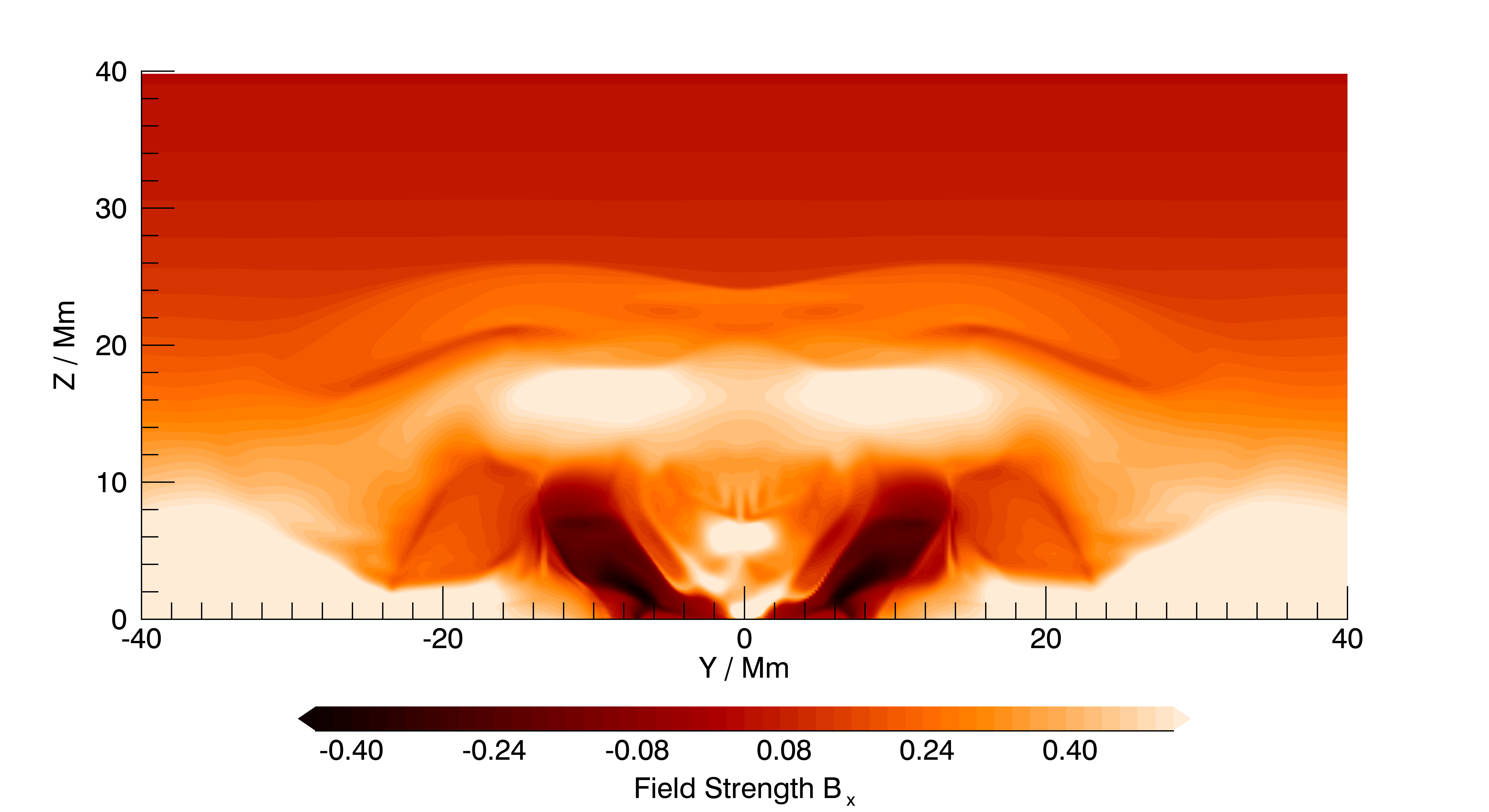}
   \hspace{0.05\linewidth}
   \includegraphics[width=0.46\linewidth]{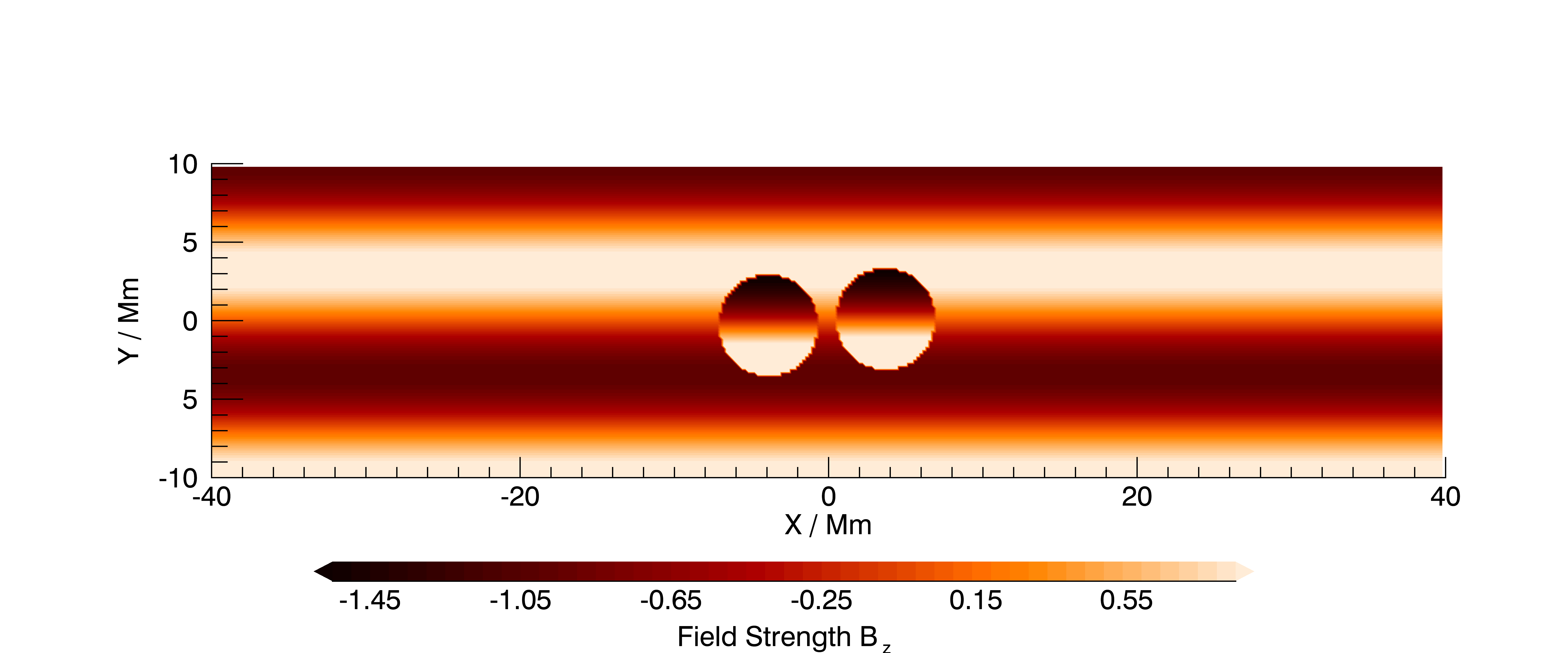}
 \endminipage\hfill
 \minipage{\textwidth}
 \centering
   \includegraphics[width=0.46\linewidth]{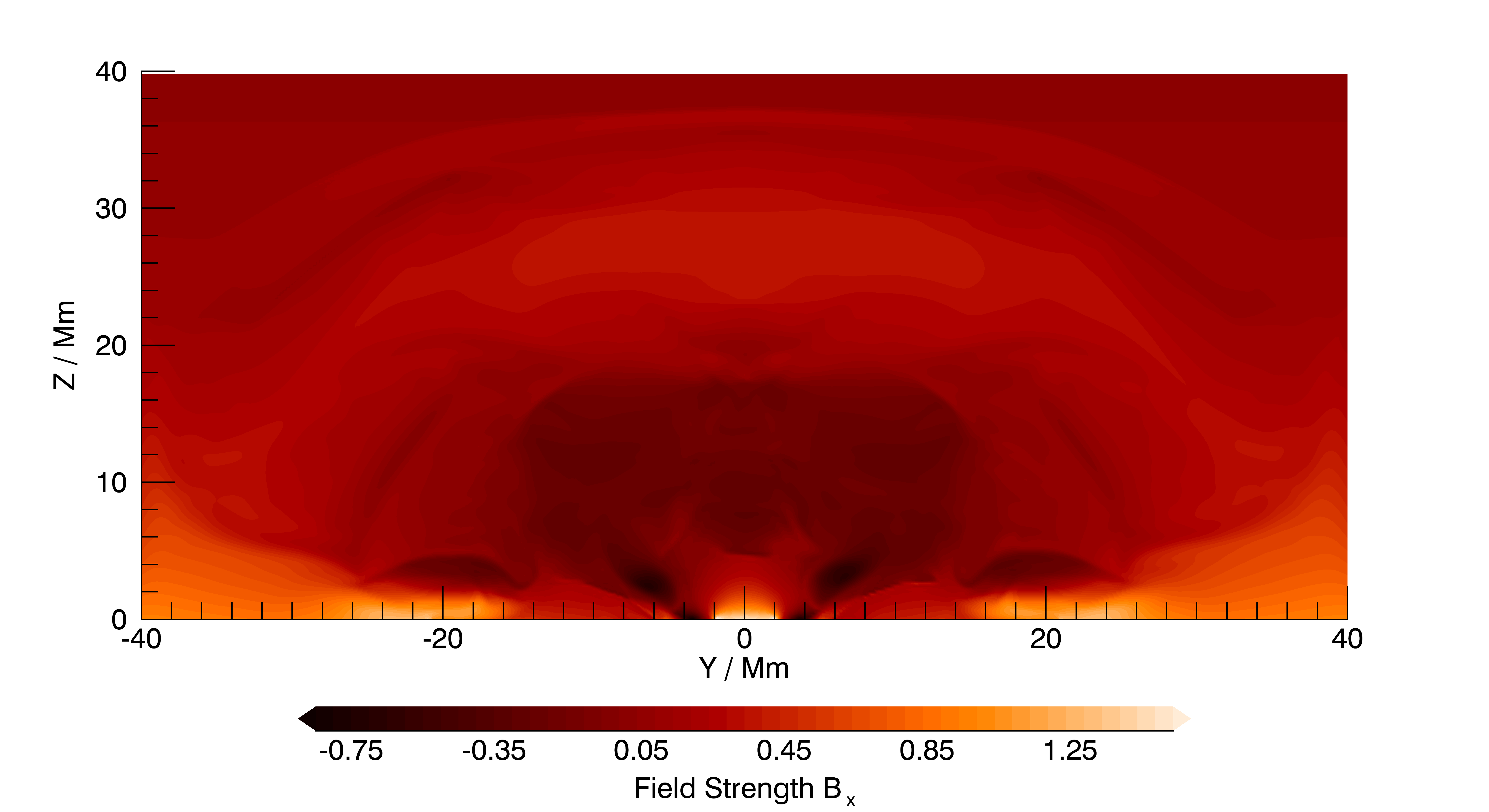}
   \hspace{0.05\linewidth}
   \includegraphics[width=0.46\linewidth]{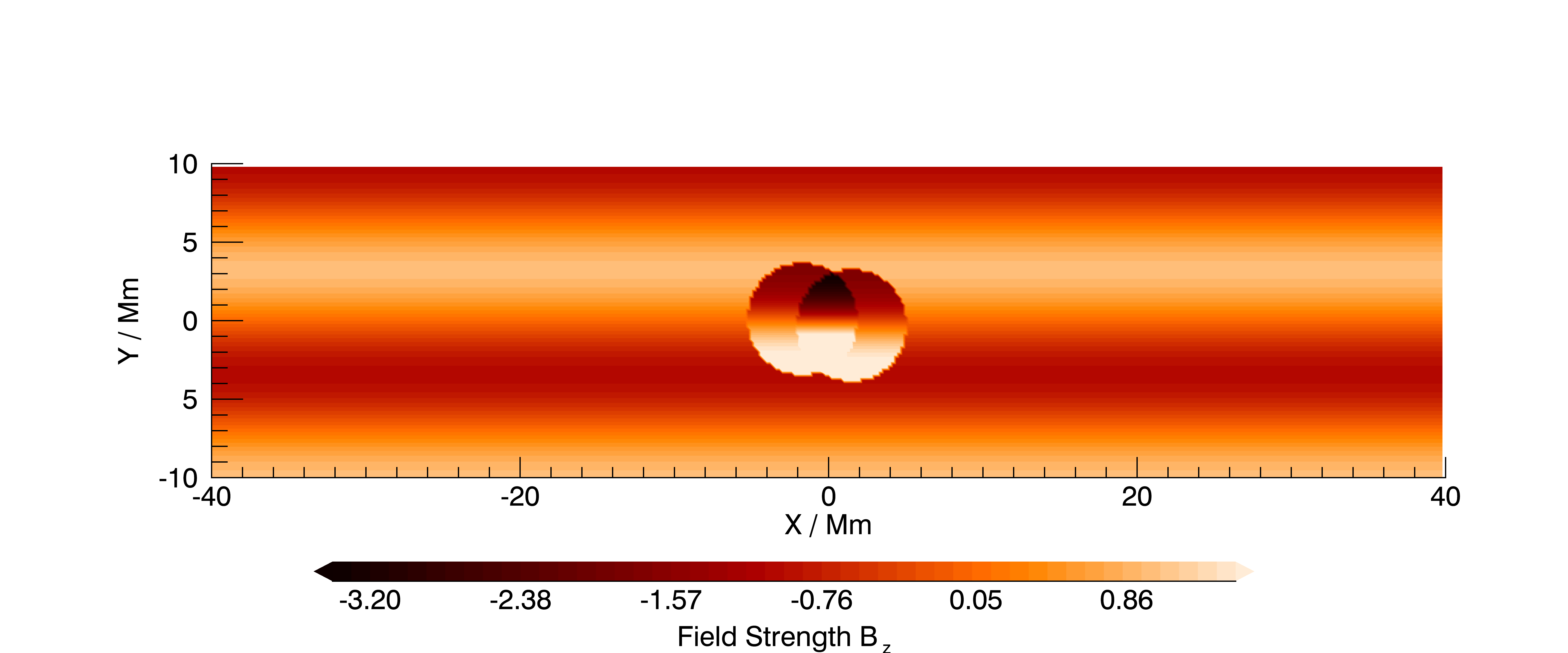}
 \endminipage\hfill
 \caption{Contour plots of (left) $B_x$ through the plane $y=0$ and (right) $B_z$ through the plane $z=0$ at times 0, 2.7, 5.4, 8.1, and 10.8 $\tau_A$ for the simulation with two colliding small-scale BRs. The magnetic reconnection can be seen occurring first at the location of each emerging field resulting in a single large flare once the BRs collide.}
 \label{fig:recon_collide}
 \end{figure*}

The aim of this study was to investigate the effects of the movement of emerging BRs on flare triggering. The results of our simulations show that the primary factor affecting the strength of the solar flares is how much of the overlying sheared field is disrupted by the emerging BRs. This is maximized when:
\begin{enumerate}
  \item The BR remains on the PIL and does not cross it.
  \item The BR remains at the preferred orientation with opposite polarity to the overlying field.
  \item The BR moves along the PIL to disrupt as much of the field as possible in a short time.
\end{enumerate}
The presence of multiple emerging BR in the same region will also increase the strength of the flare, particularly if they are both in motion. The effect of a collision will be to produce one larger flare rather than two smaller ones as seen in \cref{fig:recon_collide}. 

The results of this study are important and show that in addition to the sheared angle of the overlying field $\theta_0$ and the orientation of the overlying field $\psi_e$ the motion of the emerging BR will also have a major effect on the strength of the triggered flare. The presence of multiple small-scale BRs will also have a major effect on the flare strength particularly in the event of a collision.

\section{Acknowledgements}

D.T.and C.B. would like to acknowledge financial 
support from 2019 ISEE International Joint Research Program (PI: David 
Tsiklauri) and also would also like to cordially thank Prof. Kanya Kusano for 
his kind hospitality and stimulating atmosphere during their visits to 
ISEE, Nagoya, Japan (C.B., 6-24 May, 2019; D.T., 17-24 
August, 2019). C.B. would like to thank UK STFC DISCnet for 
financial support of his PhD studentship. This research utilized Queen 
Mary's Apocrita HPC facility, supported by QMUL Research-IT \cite{King}.
The authors would also like to thank an anonymous referee for useful suggestions that led to improvements to the manuscript.

\bibliography{flare_trigger_paper.bib}{}
\bibliographystyle{aasjournal}

\end{document}